\documentclass[final,5p,times,twocolumn]{elsarticle}
\usepackage{epsf}
\usepackage{graphicx} 
\usepackage{longtable} 
\usepackage{amssymb}
\usepackage{amsmath}
\usepackage{ulem} %% for strike-through
\usepackage[usenames]{color}
\newcommand{\be}{\begin{eqnarray}}
\newcommand{\ee}{\end{eqnarray}}
\newcommand{\non}{\nonumber \\}
\newcommand{\po}{{\rm P}}
\newcommand{\npo}{{\rm NP}}
\newcommand{\uns}{{\rm eff}}

\journal{Nuclear Physics A}

\begin{document}
\begin{frontmatter}

\author[juli]{M.~D\"oring}%\corref{cor1}\fnref{label2}
\ead{m.doering@fz-juelich.de}
\author[juli,ias]{C.~Hanhart}
\author[uga]{F.~Huang}
\author[juli,ias]{S.~Krewald} 
\author[juli,ias,bonn]{U.-G.~Mei\ss ner} 
\address[juli]{Institut f\"ur Kernphysik and J\"ulich Center for Hadron Physics, 
Forschungszentrum J\"ulich, D-52425 J\"ulich,Germany}%\fnref{label3}
\address[ias]{Institute for Advanced Simulation,
Forschungszentrum J\"ulich, D-52425 J\"ulich,Germany}
\address[uga]{Department of Physics and Astronomy, University of Georgia, Athens, Georgia 30602, USA}
\address[bonn]{Helmholtz-Institut f\"ur Strahlen- und Kernphysik (Theorie) and Bethe Center for Theoretical Physics, 
Universit\"at Bonn, Nu\ss allee 14-16, D-53115 Bonn, Germany
}

\title{\hspace{14cm}{\tiny FZJ-IKP-TH-2009-11, HISKP-TH-09/14}
\\
Analytic properties of the scattering amplitude and resonances parameters in a meson exchange model}

\begin{abstract} 
The analytic properties of scattering amplitudes provide important information. Besides the cuts, the poles and zeros on the different Riemann sheets determine the global behavior of the amplitude on the physical axis. Pole positions and residues allow for a parameterization of resonances in a well-defined way, free of assumptions for the background and energy dependence of the resonance part. This is a  necessary condition to relate resonance contributions in different reactions. In the present study, we determine the pole structure of pion-nucleon scattering in an analytic model based on meson exchange. For this, the sheet structure of the amplitude is determined. To show the precision of the resonance extraction and discuss phenomena such as resonance interference, we discuss the $S_{11}$ amplitude in greater detail.
\end{abstract}

\end{frontmatter}

\section{Introduction}
The global properties of a scattering amplitude are determined by the kinematics of the reaction, leading to branch cuts associated with the opening of reaction channels of stable or unstable particles. The thresholds of two particles or quasi-two particles being on-shell are characterized by the branch points. There is a righthand and a lefthand cut associated with $s$ channel and crossed channel processes, and there can be resonances, bound and virtual states. Resonances and virtual states are associated with poles on unphysical sheets. Thus, an analytic continuation along the various branch cuts is mandatory to access the resonance poles. 

The $\pi N\to\pi N$ transition is one of the most precisely measured reactions. It provides detailed information
about the baryon spectrum, which is presently under experimental investigation, see e.g. Ref. \cite{nstar07}.
Masses, widths, and decays of baryonic resonances
allow for tests of models of the internal structures of
the nucleon and its excited states.

Yet, the resonances in the second resonance region have to be disentangled. Thus, most of the prominent resonances listed by the PDG \cite{Amsler:2008zz} have been obtained by
partial wave analyses~\cite{Koch:1980ay,hoehlerpin,Arndt:2006bf} followed by a model dependent analysis of the partial
wave amplitudes e.g. in terms of a background and Breit-Wigner resonances 
\cite{Arndt:2006bf,Cutkosky:1979fy}. 
In the energy range between 2 and 3 GeV, presently under experimental
investigation, 
resonances start to overlap and the background may show some non-trivial structures. This situation calls for more sophisticated theoretical analyses as e.g. carried out in the partial wave analyses of Refs. \cite{Arndt:2006bf,Workman:2008iv} where poles in the
complex plane of the scattering energy are determined. 

In the case of overlapping resonances and resonances near thresholds, partial decay widths can only be extracted within models. When different reactions are analyzed simultaneously, this procedure becomes questionable. We therefore stress that the only sensible parameters that encode the resonance properties are the various pole positions and residues.

Models of the $K$ matrix type
\cite{Sarantsev:2005tg,Anisovich:2005tf,Manley:1992yb,Feuster:1997pq,Penner:2002ma,Shklyar:2004ba,Vrana:1999nt,Ceci:2006jj}
and meson exchange models
\cite{Sato:1996gk,Surya:1995ur,Schutz:1998jx,Krehl:1999km,Gasparyan:2003fp,JuliaDiaz:2007kz,Durand:2008es,Kamano:2008gr} 
provide unitary amplitudes that have been constructed in the past to access pion-nucleon scattering.
In unitarized chiral perturbation theory, resonances can be generated by the non-perturbative interaction of mesons and baryons without the need to explicitly introduce resonance propagators \cite{Dobado:1992ha,Kaiser:1995cy,Kaiser:1996js,Oller:1997ti,kaon,Oller:1998hw,Meissner:1999vr,Oller:2000fj,Inoue:2001ip,Jido:2003cb,Kolomeitsev:2003kt,Sarkar:2004jh,Doring:2005bx}. 

In this context, one has to identify observables that allow to distinguish between hadronic molecules and elementary states. Ref.~\cite{Meissner:1999vr} shows how to incorporate explicit resonance fields in a unitarized extension of chiral meson-baryon dynamics and stresses the importance of an accurate quantitative reproduction of the experimental data which is prerequisite for a determination of the pole position in the complex plane. Weinberg's method to decide on the molecular nature of bound states~\cite{weinberg} which is valid for $s$-wave states close to threshold has been generalized to resonances in Ref.~\cite{evidence}.

In some of the approaches to meson-baryon scattering, the amplitude has been analytically continued to the complex plane to extract the pole positions and parameters, see e.g Refs. \cite{Arndt:2006bf,Inoue:2001ip,Jido:2003cb,Sarkar:2004jh} and the recent work of Ref. \cite{Suzuki:2008rp} within the meson exchange framework. In the present study, we extend the amplitude of the J\"ulich model to the various Riemann sheets in the complex plane of the scattering energy $s^{1/2}\equiv z$. 

The J\"ulich model is an analytic coupled channel model based on meson exchange that respects two-body unitarity.  This model has been
developed over the past few years
\cite{Schutz:1998jx,Krehl:1999km}, with
its current form, as used in this study, given in Ref.
\cite{Gasparyan:2003fp}. For the convenience of the reader, we point out the main ideas in the following.

The coupled channel scattering equation~\cite{MuellerGroeling:1990cw,Schutz:1998jx,Krehl:1999km,Gasparyan:2003fp} is solved in the $JLS$ basis, given by
\be
&&\langle L'S'k'|T_{\mu\nu}^{IJ}|LSk\rangle=\langle L'S'k'|V_{\mu\nu}^{IJ}|LSk\rangle \non
&+&\sum_{\gamma\, L''\, S''}\int\limits_0^\infty k''^2\,dk''\langle L'S'k'|V_{\mu\gamma}^{IJ}|L''S''k''\rangle\non
&&\times\,\,\frac{1}{Z-E_{\gamma}(k'')+i\epsilon}\,\langle L''S''k''|T_{\gamma\nu}^{IJ}|LSk\rangle
\label{bse}
\ee
%\be
%T=V+VGT
%\label{bse}
%\ee 
where $J\,(L)$ is the total angular (orbital angular) momentum, $S\,(I)$ is the total spin (isospin), $k(k',\,k'')$ are the incoming(outgoing, intermediate) momenta, and $\mu,\,\nu,\,\gamma$ are channel indices. The incoming and outgoing momenta can be on- or off-shell. The integral term involves a sum over all intermediate possible quantum numbers and channels contained in the model. 

The integral term can be abbreviated as
$VGT$ where $G$ is the intermediate meson-baryon propagator of the
channels with stable particles $\pi N$ and $\eta N$, given by the fraction in Eq. (\ref{bse}). For the channels involving
quasiparticles, $\sigma N$, $\rho N$, and $\pi\Delta$, the propagator is slightly more complicated~\cite{Schutz:1998jx,Krehl:1999km}. The
pseudopotential $V$ iterated in Eq. (\ref{bse}) is constructed from an
effective interaction based on the Lagrangians of Wess and Zumino
\cite{Wess:1967jq,Meissner:1987ge}, supplemented by additional terms
\cite{Krehl:1999km,Gasparyan:2003fp} for including the $\Delta$ isobar, the
$\omega$, $\eta$, $a_0$ meson, and the $\sigma$. The exchange potentials $V$ are partial wave projected to the $J,\,L,\,S$ basis, where Eq. (\ref{bse}) is solved. Note the potentials $V$ and amplitude $T$ appear half-off-shell in the integral term of Eq. (\ref{bse}).

The exchange potentials mentioned above
contribute to the non-pole part. The pole part is given by baryonic
resonances up to $J=3/2$ that have been included in $V$ as bare $s$
channel propagators. The resonances obtain their width from the
rescattering provided by Eq. (\ref{bse}).

In this study, the partial wave amplitudes $T_{\pi N}^{IJLS}$ will be analytically continued, rather than the full amplitude or invariant amplitudes $A^\pm(s,u),\,B^\pm(s,u)$. Partial wave amplitude have a rather involved structure below the $\pi N$ threshold which is discussed in Appendix \ref{sec:app1}.
While the continuation for the channels with stable particles is straightforward, the effective $\pi\pi N$ channels require special attention. It is known~\cite{brapothree} that the quasi-two particle singularities induce novel structures in the amplitude, i.e. additional branch points in the complex plane, apart from the righthand $\pi\pi N$ cut along the physical axis. The resulting sheet structure is non-trivial and should be fully taken into account; these additional branch points induce large variations of the amplitude in their surroundings and have a large impact on pole positions and residues. For the Roper channel, the role of additional sheets has been discussed in Ref. \cite{Cutkosky:1990zh}. 

In some studies, the properties of channels with unstable particles have been modeled by folding the invariant mass distribution of the unstable particle with the stable particle propagator as e.g. in Refs. \cite{Gamermann:2007fi,Albaladejo:2008qa}. We have developed related techniques based on a Lehmann representation for the righthand cut. While the discontinuity can be expressed in this form, an explicit proof of analyticity of the obtained continuation is difficult within our framework.

For the J\"ulich model, the various sheets are accessed through contour deformation of the momentum integration. A similar method has been recently applied for another model of the meson exchange type~\cite{Suzuki:2008rp}.

Another issue of relevance in this context is 
the question, if it is possible to remove
the hadronic contributions from observables in
a model independent way to allow access to 
quantities that can be identified with those
calculated from the quark model~\cite{leenew,svarcundress} --- see
also Ref.~\cite{nstar} for a recent discussion
of the subject. Such an analysis assumes
that a clean cut separation of pole and non-pole 
parts is possible. 

In a recent study~\cite{Doring:2009bi} this question has been addressed. The conclusion, drawn from the behavior of the $P_{33}$ partial wave with the $\Delta(1232)$ resonance, was that the residues provide a meaningful expansion parameter of the resonance amplitude, while dressed and bare vertices depend on the regularization scheme and the model dependent decomposition of the amplitude into pole and non-pole contribution according to $T=T^\po+T^\npo$. This separation is widely used in the literature, see e.g. Refs. \cite{Afnan:1980hp,Matsuyama:2006rp}. Second, the non-pole part itself can contain poles, that interact with the pole part in a non-trivial way, as discussed in Ref. \cite{Doring:2009bi}. In the present study, we test the $D_{13}$ and other partial waves in the light of these questions and come to similar conclusions.

After all, especially for the comparison of different experiments, the poles and residues of the S-matrix 
are the relevant quantities. The J\"ulich model is derived within a field theoretical approach from Lagrangians obeying chiral constraints. It provides a precise description of the partial wave amplitudes. Furthermore, other than in most $K$ matrix approaches, analyticity is respected. These ingredients qualify for a reliable extraction of resonance properties.

This paper is organized as follows.
In Sec. \ref{sec:two-particle}, the well-known analytic structure of the propagator of two stable particles is discussed. The more complicated properties of quasi-twoparticle intermediate states are discussed in Sec. \ref{sec:threeparticle}. Results for pole positions, residues, zeros of the amplitude and transition strengths are provided in Sec. \ref{sec:results}.

%%%%%%%%%%%%%%%%%%%%%%%%%%%%%%%%%%%%%%%%%%%%%%%%%%%%%%%%%%%%%
%%%%%%%%%%%%%%%%%%%%%%%%%%%%%%%%%%%%%%%%%%%%%%%%%%%%%%%%%%%%%

\section{Analytic structure of the scattering amplitude}
\label{sec:anal}
\subsection{The propagator of two stable particles}
\label{sec:two-particle}
The analytic continuation of the scattering amplitude will be discussed in several steps. In this section, the well-known structure for the propagator of two stable particles is presented. The connection between analytic continuation and contour deformation of the momentum integration will be pointed out. The contour deformation is the key to the analytic continuation of the effective $\pi\pi N$ channels $\sigma N$, $\rho N$, and $\pi\Delta$, discussed in Sec. \ref{sec:threeparticle}.

To simplify the discussion, we choose pion-pion scattering in the $L=I=0$ channel. Here, the scattering problem is described as a one-channel amplitude with a  propagator of two stable pions and one explicit resonance, the $\sigma(600)\, (J^P=0^+)$ meson. In the J\"ulich model, this amplitude appears in the construction of the $\sigma N$ propagator and also in the $\sigma$ $t$-channel exchange. The example will serve to demonstrate the analytic properties of the propagator of two stable particles. The following discussion applies qualitatively also to the $\pi N$ and $\eta N$ channels of the J\"ulich model, as well as to other $\pi\pi$ and $\pi N$ channels, namely the $\rho(770)$ in the $\rho N$ propagator and the $\Delta(1232)$ in the $\pi\Delta$ propagator.

The $\sigma$ can be described with a dressed TOPT (time-ordered perturbation theory) propagator according to 
\be
G_\sigma(z,k)=\frac{1}{z-\sqrt{k^2+(m_\sigma^0)^2}-\Pi_\sigma(z',k)}
\label{sigprop}
\ee
where $m_\sigma^0=900$ MeV is the bare $\sigma$ mass, $k\equiv|\vec{k}|$ is the c.m. momentum, i.e., $k=0$ in the $\sigma$ c.m. frame, $z'$ is the energy boosted to the $\pi\pi$ c.m.
frame ($z'=z$ at $k=0$ is the only case needed in this section), and $\Pi_\sigma$ is the $\sigma$ self-energy given by the loop
\be
\Pi_\sigma(z,k)=\int\limits_0^\infty q^2dq\,\frac{(v^{\sigma\pi\pi}(q,k))^2}{z-2\sqrt{q^2+m_\pi^2}+i\epsilon}
\label{sigself}
\ee
where $v^{\sigma\pi\pi}(q,k)$ is given in Eq. (3.10) of Ref. \cite{Schutz:1998jx}. For the $\rho$ and $\Delta$ self energies, the vertices are given by Eq. (20) of Ref. \cite{Krehl:1999km} and in Ref. \cite{Schutz:1998jx}, respectively. 

\begin{figure}
\includegraphics[width=0.4\textwidth]{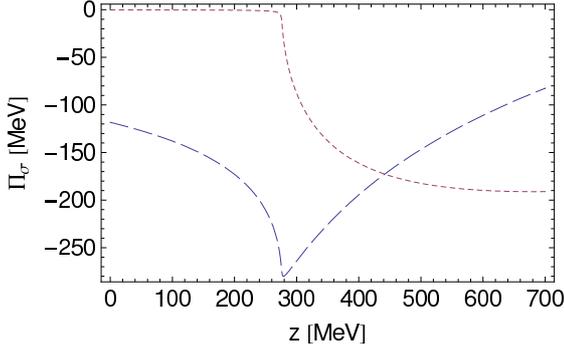}
\caption{Real (dashed) and imaginary (dotted) parts of the $\sigma$ self-energy at $k=0$ MeV.}
\label{fig:siself}
\end{figure}

\begin{figure}
\includegraphics[width=0.48\textwidth]{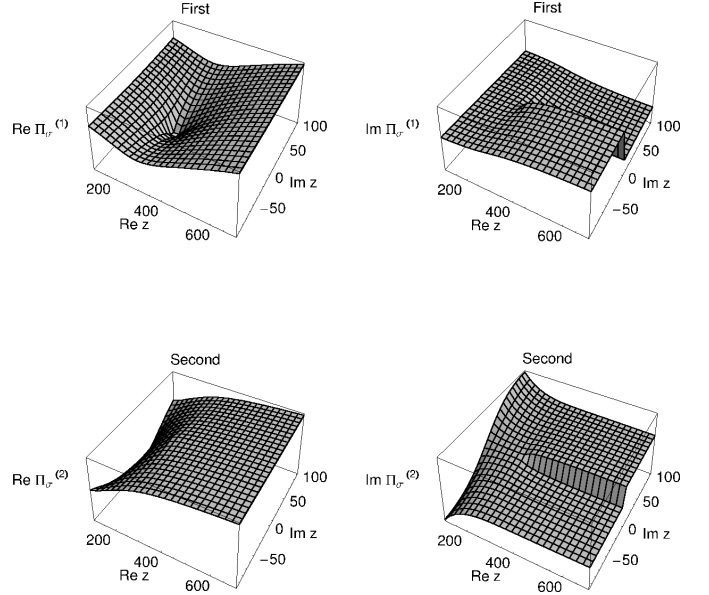}
\caption{The two Riemann sheets of the $\sigma$ self-energy [arb. units] as a function of $z$ [MeV]. The left column shows the real and the right column the imaginary part of $\Pi_\sigma$.}
\label{fig:sisheets}
\end{figure}

The $\sigma$ self-energy for real energies $z$ is shown in Fig. \ref{fig:siself}.
The imaginary part becomes finite and negative at the two-pion threshold while the real part shows a cusp at this point. 
To understand the analytic properties of the $\sigma$ self-energy, consider complex values of $z$ in Eq. (\ref{sigself}), evaluating the integral along a straight path from zero to infinity along the real $q$ axis. The self-energy as a complex function of $z$ is shown in the first line of Fig. \ref{fig:sisheets}. 
The result is called the first sheet of the self-energy. There is a righthand cut extending along the positive real $z$ axis starting at the energy where both pions can go on-shell, i.e. $z_{\rm thresh}=2\,m_\pi$. The cut is at $z-i\epsilon$, i.e. $\Pi_\sigma$ on the real $z$ axis is analytically connected to $\Pi_\sigma$ in the upper half plane. 

Along the righthand cut, the amplitude can be analytically continued. For this, the standard procedure is to calculate the imaginary part at the real axis by evaluating Eq. (\ref{sigself}) with the $\delta$ function $\delta(z-2E_q)$ resulting in
\be
{\rm Im}\,\Pi_\sigma=-\,\frac{\pi\,q_{\rm{on}}^>\,E^{(1)}_{\rm{on}}\,E^{(2)}_{\rm{on}}}{z}\,v^2(q_{\rm{on}}^>,k)
\label{impartstable}
\ee
where $q_{\rm{on}}$ is the on-shell three-momentum in the c.m. frame of the stable particles 1 and 2 (in this case two pions) and $E_{\rm{on}}$ is the on-shell energy of the particles. As the quantity 
\be
q_{\rm{on}}=\frac{1}{2z}\sqrt{(z^2-(m_1-m_2)^2)(z^2-(m_1+m_2)^2)}
\label{onstan}
\ee
is two-valued itself, we need to distinguish the two Riemann sheets of $q_{\rm{on}}$ uniquely according to
\be
q_{\rm{on}}^>&=&
\begin{cases}
-q_{\rm{on}}	&	\text{if Im $q_{\rm{on}}<0$}\\
\,\,q_{\rm{on}}	&	\text{else}
\end{cases}\non
q_{\rm{on}}^<&=&\,\,-q_{\rm{on}}^>.
\label{sq_pres}
\ee
With this definition, $q_{\rm{on}}^>$ is real and positive on the real axis above threshold; it has a cut along the real axis at $z-i\epsilon$ above $z_{\rm thresh}$, where the branch point is located, and is analytic otherwise. 

\begin{figure}
\begin{center}
\includegraphics[width=0.46\textwidth]{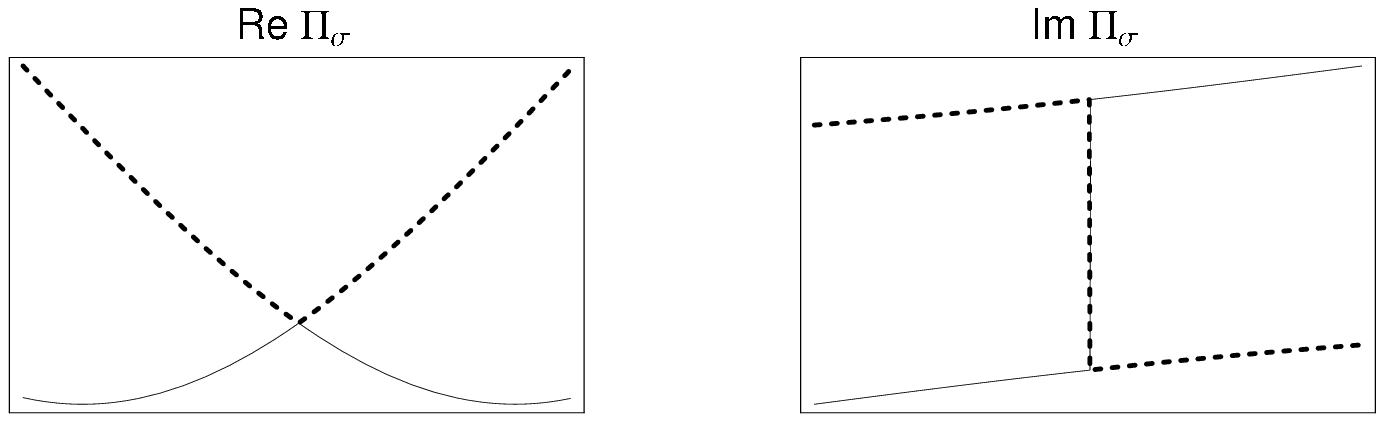}
\includegraphics[width=0.48\textwidth]{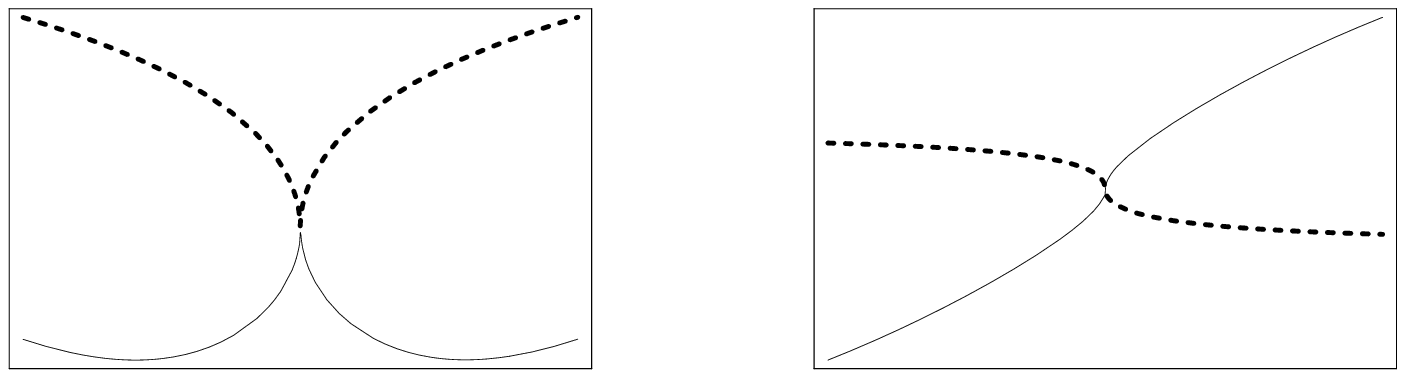}
\includegraphics[width=0.476\textwidth]{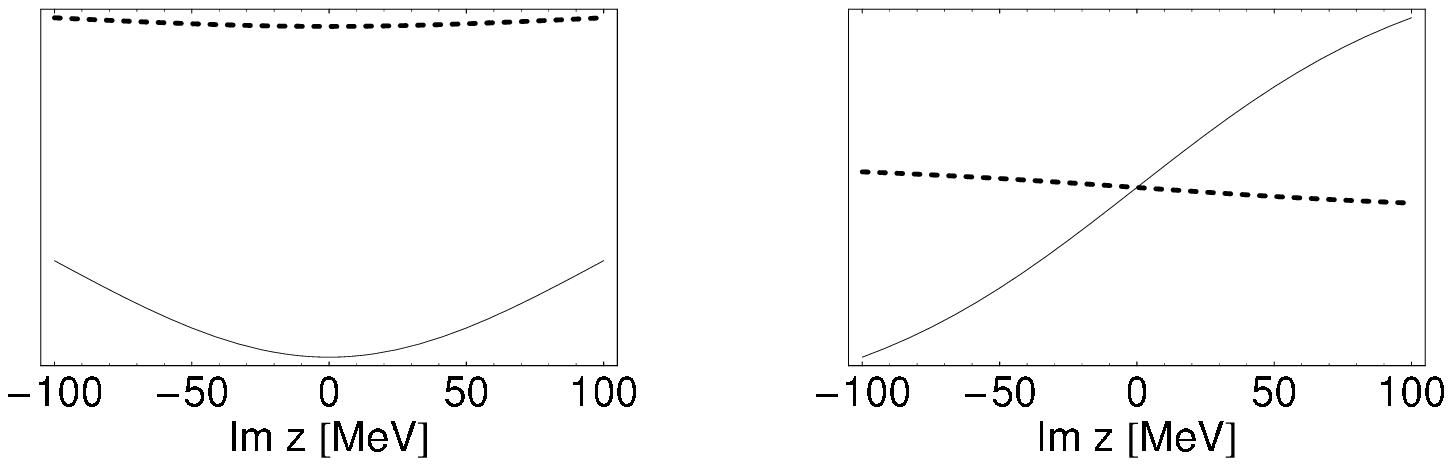}
\end{center}
\caption{First (dashed lines) and second sheet (solid lines) of the $\sigma$ self-energy $\Pi_\sigma$ [arb. units]. The upper row shows a slide for ${\rm Re}\,z$ above the $\pi\pi$ threshold (note the discontinuity at Im $z=0$), the middle row for ${\rm Re}\,z=2m_\pi$ and the lower row for ${\rm Re}\,z$ below the threshold.}
\label{fig:slides_pisig}
\end{figure}

The analytic continuation to the second sheet is given by
\be
\Pi_\sigma^{(2)}=\Pi_\sigma+\frac{2\pi i\,q_{\rm{on}}^>\,E^{(1)}_{\rm{on}}\,E^{(2)}_{\rm{on}}}{z}\,v^2(q_{\rm{on}}^>,k),
\label{prescristable}
\ee
i.e., by subtracting twice the imaginary part of Eq. (\ref{impartstable}). In Fig. \ref{fig:slides_pisig} we show the two sheets $\Pi^{(1)}_\sigma\equiv \Pi_\sigma$ (solid lines) and $\Pi^{(2)}_\sigma$ (dashed lines) as a function of ${\rm Im}\,z$ for fixed ${\rm Re}\,(z)$.
In Fig. \ref{fig:sisheets} the same situation is shown in a three-dimensional plot. 

The two sheets are analytically connected along the righthand cut. Note that below threshold the second sheet is not directly connected to the physical axis but only via paths around the branch point at $z_{\rm thresh}$. Above threshold, there is a direct connection from the second sheet in the lower half plane to the physical axis.

In the following we sketch the proof of analyticity of the prescription from Eq. (\ref{prescristable}). Although this proof is trivial it will help us understand the procedure of the analytic continuation for unstable particles in Sec. \ref{sec:threeparticle}. 

\begin{figure}
\includegraphics[width=0.38\textwidth]{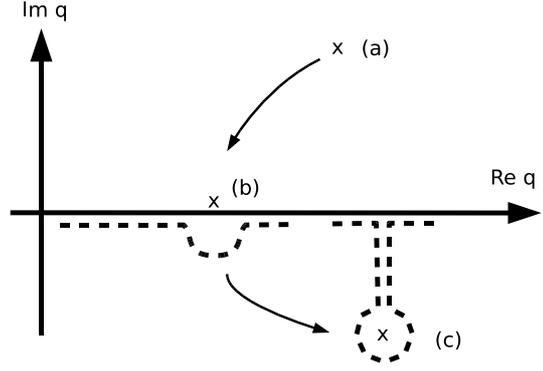}
\caption{Integration contours (dashed lines) for the loop of two stable particles ($\pi\pi$). For different $z$, the poles are marked with a cross. The arrows indicate the changes of the pole positions for the cases ${\rm Im}\, z>0$, $=0$, $<0$.}
\label{fig:analy_scheme}
\end{figure}

Consider the pole of the integrand from Eq. (\ref{sigself}) with respect to $q$ in the right $q$ half plane. Its position, for different $z$ values, is indicated in Fig. \ref{fig:analy_scheme} with crosses. 
For $z$ values with positive imaginary part --shown as case (a)-- the pole is in the upper $q$ plane and the $q$ integration can be carried out on a straight path from zero to infinity. For real $z$ --this is the relevant case to evaluate observables-- the pole is on the real axis [cf. case (b)] and one can deform the integration contour with a half circle as shown in Fig. \ref{fig:analy_scheme}. The evaluation of the infinitesimal half circle returns the imaginary part of the self-energy. Once ${\rm Im}\,z <0$, the pole is in the lower half plane. This crossing of the pole over the integration contour corresponds to the appearance of the right-hand cut. Yet, one can obtain an analytic continuation for negative ${\rm Im}\,z$ by defining a deformed path shown as case (c) in Fig. \ref{fig:analy_scheme}. 

From the construction of the path in case (c) it becomes clear that this continuation is indeed analytic. From the form of the path it is also clear that the result of the integration corresponds to an integration along a straight path on the real axis (corresponding to the first sheet), plus the integral taken on the full infinitesimal circle, that returns the residue of the integration. The contributions from the vertical parts of the path (c) in Fig. \ref{fig:analy_scheme} cancel. The final result is then exactly given by Eq. (\ref{prescristable}), as an explicit evaluation of the residue shows. 

To finish this discussion, let us again consider the poles in the $q$-plane. The pole at $q_p$ in the right half plane is simple; it is accompanied by another simple pole at $-q_p$. However, for the energy $z_{b_1}=m_1+m_2$ there is only one double pole in the momentum plane, situated at $q=0$. The energy $z_{b_1}$, however, is threshold and branch point of the Riemann surface. 

Concluding, a branch point for integrals of the type of Eq. (\ref{sigself}) is induced whenever the two poles in the $q$-plane coincide and form a double pole. We will also encounter this situation in the more complicated case of the effective $\pi\pi N$ propagators, where the classification of the Riemann surface in terms of branch points is more complicated.

Eq. (\ref{prescristable}) gives the analytic continuation of the self-energy of two stable pions. In the J\"ulich model, there are the two channels with stable particles $\pi N$ and $\eta N$. The analytic continuation of the amplitude with respect to a channel (mn) given by a stable meson $m$ and a stable baryon $n$ is closely related to the second term in Eq. (\ref{prescristable}), i.e. the discontinuity along the righthand cut.
In particular, we obtain the scattering equation for the amplitude $T^{(2)}$ on the second sheet, 
\begin{multline}
\langle q_{cd}|T^{(2)}-V|q_{ab}\rangle=\\
\delta G+\int dq_{mn}\,q_{mn}^2\frac{\langle q_{cd}|V|q_{mn}\rangle\langle q_{mn}|T^{(2)}|q_{ab}\rangle}{z-E_{mn}+i\epsilon}\nonumber
\end{multline}
\begin{multline}
\delta G=\frac{2\pi i\,q_{\rm{on}}^>(mn)\,E_{m}^{\rm{on}}\,E_n^{\rm{on}}}{z}\\
\times\langle q_{cd}|V|q_{\rm{on}}^>(mn)\rangle\langle q_{\rm{on}}^>(mn)|T^{(2)}|q_{ab}\rangle
\label{analpin}
\end{multline}
with $E_{mn}=E_m+E_n$ and $q_{\rm{on}}^>$ from Eq. (\ref{sq_pres}). Indices of quantum numbers and the angle integration have been suppressed in Eq. (\ref{analpin}). Eq. (\ref{analpin}) has a similar form as the scattering equation on the first sheet from Eq. (\ref{bse}), except for the additional term $\delta G$, which is on the real axis given by the discontinuity of the two particle propagator $1/(z-E_{mn})$. In Eq. (\ref{analpin}), the indices (ab), (cd), and (mn) indicate the incoming, outgoing, and intermediate channels, respectively.

Note that the matrix elements $V$ and $T^{(2)}$ in the expression for $\delta G$ appear in on-shell kinematics which allows for an implementation of the term at the on-shell point of the Haftel-Tabakin scheme \cite{hafteltabakin} that is used to solve the scattering equation.

One can now also search for poles of the $\sigma(600)$, $\rho(770)$, and $\Delta(1232)$ as they appear in the effective $\pi\pi N$ propagators of the J\"ulich model. Simple but analytic models have been developed to obtain a good empirical parameterization of the $\pi\pi$ and $\pi N$ amplitude which are needed as input just for a parameterization of the effective $\pi\pi N$ channels. This purpose is well fulfilled by these models. 

For the pole search, we analytically continue the propagator from Eq. (\ref{sigprop}), with $\Pi^{(2)}_\sigma$ from Eq. (\ref{prescristable}), according to 
\be
G_\sigma^{(2)}(z_\sigma)=\frac{1}{z_\sigma-m_\sigma^0-\Pi^{(2)}_\sigma(z_\sigma,0)}
\label{gsigzwei}
\ee
as $k=0$ in the $\sigma$ c.m. frame. A pole on the second sheet of the $\pi\pi$ scattering amplitude corresponds to a pole of $G_\sigma^{(2)}(z_\sigma)$. We find the pole of the $\sigma$ at $z^0_\sigma=875-232\,i$ MeV. The $\sigma$ pole is situated relatively high in energy compared to the high precision determination in Ref. \cite{Caprini:2005zr} of $z^0_\sigma=441-272\,i$ MeV. At this point the model could be improved by developing a scheme with derivative coupling that allows for a $\sigma$ pole at lower energies. In any case, a satisfying parameterization of the $\pi\pi$ scattering amplitude in the $\sigma$ channel is obtained. In Fig. 10 of Ref. \cite{Krehl:1999km} the quality of the fit is shown. There are some minor deviations, probably due to the displaced $\sigma$ pole, but for the present purpose, this accuracy is sufficient to provide a realistic description of the three body phase space, starting at $z=2m_\pi+m_N$ up to the maximum energies considered of around $z\sim 1.9$ GeV. In the complex plane, a change in the pole position of the $\sigma$ would lead to a change of the position of the branch point (cf. Eq. (\ref{polebra})); however, as the pole is quite far in the complex plane, no major changes from a modified $\sigma$ pole are expected in the region where poles of baryonic resonances are searched for ($\Gamma/2\le 150\,i$ MeV).

For the $\rho$, a similar analysis can be made. There are no poles on the first sheet and one pole on the second sheet at $z_\rho^0=763-64\,i$ MeV. This corresponds to a $\rho$ width of $\Gamma=128$ MeV which is slightly smaller than the standard value of $\Gamma=150$ MeV \cite{Amsler:2008zz}. 

For the $\Delta$, the pole lies at $z_\Delta^0=1211-37\,i$ MeV, i.e. the $\Delta$ has a width of $\Gamma=74$ MeV which is smaller than standard values of 110-120 MeV~\cite{Amsler:2008zz}. The pole position of the $\Delta(1232)$ in the $\pi\Delta$ propagator is not identical to the position of the $\Delta(1232)$ in the $\pi N$ $s$-channel exchange, that will be determined later (cf. Ref. \cite{Doring:2009bi} or Table \ref{tab:reso}). The consistency of the model could be improved at this point, i.e. consistent pole position for both cases. However, the discrepancies are minor. As we have stressed before, at this point a high precision fit of the amplitude is not required; a reliable parameterization of the input for the effective $\pi\pi N$ channels is sufficient.

Thus, the $\pi\pi$ and $\pi N$ phase shifts are sufficiently well described (cf. Fig. 7 of Ref.~\cite{Krehl:1999km}, Figs. 9 and 10 of Ref.~\cite{Schutz:1998jx}) by the $\sigma,\,\rho$, and $\Delta$ to allow for a realistic description of the effective $\pi\pi N$ channels as they appear in $\pi N$ scattering, the principal objective of this study.

%%%%%%%%%%%%%%%%%%%%%%%%%%%%%%%%%%%%%%%%%%%%%%%%%%%%%%%%%%%%%%%%%%%%%%%

\subsection{Propagator with unstable particles}
\label{sec:threeparticle}
The analytic continuation for the effective $\pi\pi N$ channels $\sigma N$, $\rho N$, and $\pi \Delta$ is different from the channels $\pi N$ and $\eta N$ discussed in the previous section. There, we have seen that a deformation of the integration contour as shown in Fig. \ref{fig:analy_scheme} leads to an analytic continuation. In practical terms, one can simply add the discontinuity using the $\delta$ function which is equivalent. For the effective $\pi\pi N$ channels, the key to the analytic continuation is the contour deformation. 

For the discussion, we focus on the $\sigma N$ channel. The $\rho N$
and $\pi \Delta$ channels can be treated analogously as discussed at
the end of Sec. \ref{sec:formalpath}.
The additional complication that arises here is that the unstable
particles appear in a  moving frame. Within TOPT, the three dimensional
formalism used here, the $\sigma$ self--energy evaluated for a finite sigma
three momentum $k$ reads
\be
\tilde{\Pi}_\sigma(z,k)=\int\limits_0^\infty q^2dq\, \int\limits_{-1}^1 \frac{dx}{2}\, \frac{(v^{\sigma\pi\pi}(q,k))^2}{z-
\omega_\pi^+-\omega_\pi^-+i\epsilon} \ ,
\label{sigself_boost}
\ee
with $\omega_\pi^\pm=\sqrt{(\vec q\pm \vec k/2)^2+m_\pi^2}$ and $x$ for the angle enclosed by
$\vec q$ and $\vec k$. This expression
reduces to the one given in Eq.~(\ref{sigself}) for $k=0$.
Note, $\tilde{\Pi}_\sigma$ is a function of only the modulus of $\vec k$ and not its direction
as a result of the $x$--integration. It is instructive to expand
the denominator of the self--energy for small values of $k$. We then get
$$
z-\omega_\pi^+-\omega_\pi^- = z - 2\omega_\pi - k^2/(4\omega_\pi) + {\mathcal O}(k^4/\omega_\pi^3)  \ ,
$$
with $\omega_\pi=\sqrt{q^2+m_\pi^2}$. Thus, through the boost momentum 
$k$ the energy available for the $\sigma$ gets reduced
 by $k^2/(4\omega)$, the
kinetic energy of the two pion system. We may therefore parameterize
 the $\sigma N$ propagator as~\cite{Schutz:1998jx}
 \be &&g_{\sigma N}(z,k)
=\non
&&\frac{1}{z-\sqrt{m_N^2+k^2}-\sqrt{(m_\sigma^0)^2+k^2}-\Pi_\sigma(z_\sigma(z,k),k)},\non
&&G_{\sigma N}(z)=\int\limits_0^\infty dk\,k^2\, F(k)\,g_{\sigma
  N}(z,k),\non
&&z_\sigma(z,k)=z+m_\sigma^0-\sqrt{k^2+(m_\sigma^0)^2}-\sqrt{k^2+m_N^2}\non
\label{signpro}
\ee with the self energy $\Pi_\sigma$ from Eq. (\ref{sigself});
$m_\sigma^0$ is the bare $\sigma$ mass and $F$ is a regulator we
introduce for the discussion of this section and which is absent in
the J\"ulich model [cf. Sec. \ref{sec:implement}]. The term $g_{\sigma
  N}$ is given by the TOPT $\sigma N$ propagator that includes the
self energy of the $\sigma$ meson. 
The energy parameter  $z_\sigma$ implies that the
energy available for the $\pi\pi$ subsystem is reduced
compared to the total energy not only by the kinetic energy of the $\sigma$, but
also by the energy of the nucleon propagating simultaneously.
We checked numerically that treating $\Pi_\sigma$ as in 
Eq.~(\ref{sigself}) vs. keeping the full dependence of $\tilde{\Pi}_\sigma$
of Eq.~(\ref{sigself_boost}) has only a very small impact
on observables. Since using Eq.~(\ref{signpro}) saves an angular
integration we will use this expression in what follows.

\begin{figure}
\begin{center}
\includegraphics[width=0.2\textwidth]{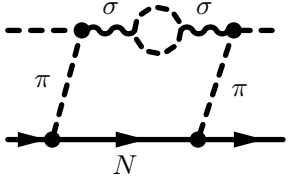}
\end{center}
\caption{Example of an intermediate $\sigma\,N$ propagator in the rescattering provided by the scattering equation (\ref{bse}).}
\label{fig:examplesn}
\end{figure}

In Eq. (\ref{signpro}), we have explicitly included the integration over $k$, the loop momentum of the $\sigma N$ loop. As an example of such a loop, we show in Fig. \ref{fig:examplesn} two pion exchanges.
The exchange processes are $k$ and angle dependent; they induce three-body cuts and additional analytic structures, on top of the structure given by $g_{\sigma N}$. This issue will be further discussed in  Sec. \ref{sec:implement} where also the implementation of the analytically continued propagator into the full J\"ulich model is given. See also Appendix \ref{sec:app1} for a discussion of the short nucleon, circular and other cuts.

\begin{figure}
\includegraphics[width=0.40\textwidth]{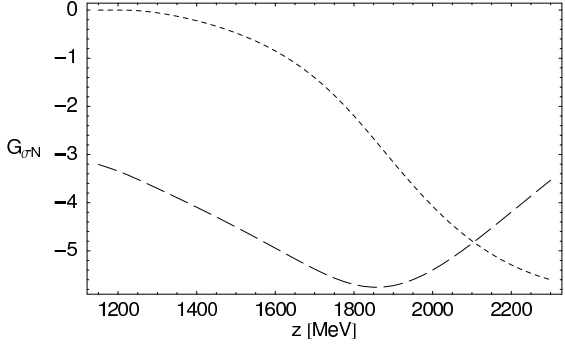}
\caption{Real (dashed) and imaginary (dotted) parts of the $\sigma N$ propagator $G_{\sigma N}$ for real energies $z$ [arb. units].}
\label{fig:snprop}
\end{figure}

In the following, the analytic structure of $G_{\sigma N}$ as defined in Eq. (\ref{signpro}) is determined. For real energies $z$, the propagator $G_{\sigma N}$ is shown in Fig. \ref{fig:snprop}.
The real and imaginary parts can be compared to the case of stable particles as shown in Fig. \ref{fig:siself}. One observes similar structures, which are, however, smeared in energy due to the finite width of the $\sigma$, or equivalently, a finite $\Pi_\sigma$ in Eq. (\ref{signpro}). In particular, the influence of the ``threshold'' at $z\sim m_\sigma^0+m_N\sim 1.9$ GeV is still visible and reminds one of the corresponding structure in case of stable particles.

The $\pi\pi$ self-energy $\Pi_\sigma$ in Eq. (\ref{signpro}) has a well-known righthand cut along the real $z_\sigma$ axis as derived in Sec. \ref{sec:two-particle}. It is clear that this induces also a cut in the full propagator $G_{\sigma N}$. 

In the case of the propagator of two pions the discontinuity arises from the position of the pole with respect to the integration contour, as discussed following Fig. \ref{fig:analy_scheme}. In particular, for real $z$, the pole in the $q$-integration lies on the real axis. In the present case, the pole lies, for real $z$, far in the complex plane [c.f. Eq. (\ref{signpro})], and the cut of $G_{\sigma N}$ on the real $z$ axis, starting at $z=2m_\pi+m_N$, is entirely induced by the cut of $\Pi_\sigma$ itself. 

Thus, for an analytic continuation of $G_{\sigma N}$, one first has to analytically continue $\Pi_\sigma$. For this, we consider the $k$ integration in Eq. (\ref{signpro}). The integration contour in the complex $k$ plane can be deformed as long as the limits $(0,\infty)$ are unchanged. Second, the deformation must be along analytic regions of the integrand and must not cross poles, branch points, or cuts. This guarantees that the result of the integration along the deformed contour is unchanged. 

\begin{figure}
\includegraphics[width=0.35\textwidth]{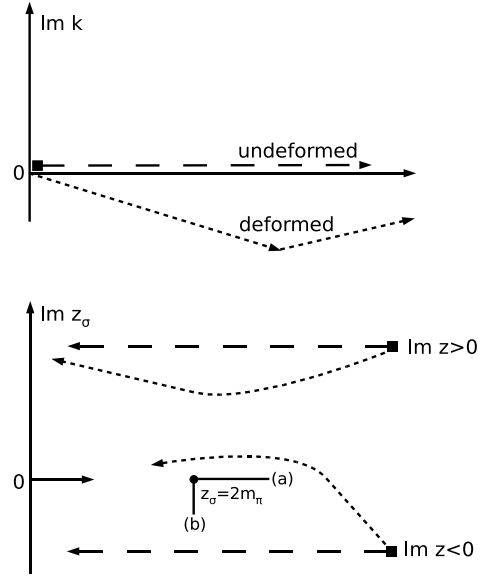}
\caption{In the $k$ plane (upper figure), the path of the $k$-integration (long dashed line) can be deformed (short dashed line). In the $z_\sigma$ plane, this corresponds to different paths for different values of $z$ (lower figure). In the $z_\sigma$ plane, there is a branch point at $z_\sigma=2m_\pi$. The corresponding cut can be chosen in different directions, e.g. (a) or (b).}
\label{fig:defo1}
\end{figure}

To ensure these conditions, we have to check which are the $z_\sigma$ regions in $\Pi_\sigma$ that correspond to the contour deformation in the $k$ plane. A path in the $k$ plane corresponds to a path in the $z_\sigma$ plane as given by the transformation $z_\sigma(z,k)$ from Eq. (\ref{signpro}). We show in Fig. \ref{fig:defo1} schematically several deformed paths in the $k$- and $z_\sigma$-plane.

First, consider the undeformed path for a value of ${\rm Im}\, z\equiv{\rm Im}\, z_\sigma>0$, i.e. a straight path in the $k$ plane from zero to infinity. This is indicated with the long dashed lines in Fig. \ref{fig:defo1}. The integral along the deformed path (short dashed lines), given that the endpoint is the same as for the undeformed path, will return the same result because the deformation is over a region in $z_\sigma$ where $\Pi_\sigma$ is analytic. 

Next, consider the case  ${\rm Im}\, z\equiv{\rm Im}\, z_\sigma<0$. The righthand cut of $\Pi_\sigma$ is along the real $z_\sigma$ axes, indicated as direction (a) in Fig. \ref{fig:defo1}. The righthand cut begins at the branch point $z_\sigma=2m_\pi$. Thus, the integration along the undeformed path induces a cut in the full $G_{\sigma N}$ once ${\rm Im}\, z$ changes sign. 

The analytic continuation of $G_{\sigma N}$ along this cut, for ${\rm Im}\, z<0$, is obtained as follows: If ${\rm Im}\, z$ changes from positive to negative values, the integration contour is deformed as indicated with the short dashed line in the $z_\sigma$ plane in Fig. \ref{fig:defo1} (case ${\rm Im}\, z<0$). However, this is possible only if simultaneously the cut of $\Pi_\sigma$ is moved from direction (a) to direction (b). This ensures that the integrand is always analytic. The continuation obtained through this contour deformation is the unique analytic continuation of $G_{\sigma N}$.

\begin{figure}
\includegraphics[width=0.45\textwidth]{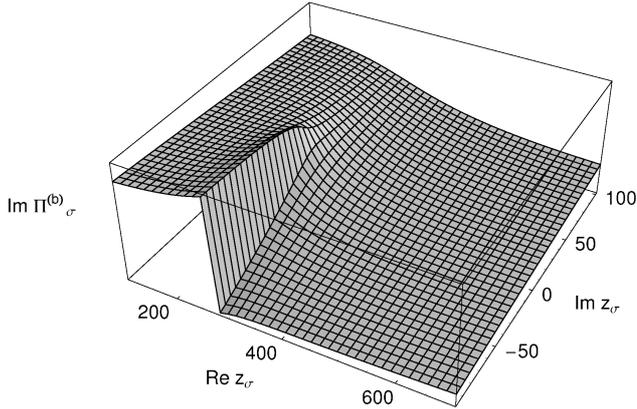}
\caption{Imaginary part of the self-energy $\Pi_\sigma^{(b)}$ [arb. units] as a function of $z_\sigma$ [MeV]. The cut is put in the negative ${\rm Im}\, z_\sigma$ direction. With this choice, the self-energy is analytic on the real axis above the branch point.}
\label{fig:new_pise}
\end{figure}

The change of direction of the cut from (a) to (b) redefines the self-energy according to 
\be
\Pi_\sigma^{(b)}=
\begin{cases}
\Pi_\sigma^{(2)}&	\text{if Im $z_\sigma<0$ and Re $z_\sigma>2m_\pi$}\\
\Pi_\sigma	&	\text{else}
\end{cases}
\label{bsheet}
\ee
where $\Pi_\sigma$ and $\Pi_\sigma^{(2)}$ are given in Eqs. (\ref{sigself}) and (\ref{prescristable}), respectively. The self-energy $\Pi_\sigma^{(b)}$ is shown in Fig. \ref{fig:new_pise}.
Indeed, the cut is now along direction (b) and $\Pi_\sigma^{(b)}$ is analytic on the real axis above the branch point.

Second, as shown above, the integration path has to be deformed; for simplicity, we have decomposed the path into straight pieces in the $k$ plane in such a way that one intermediate edge point $k_{\rm int}$ corresponds to $z_\sigma(k_{\rm int})=2\,m_\pi+100\,i$ MeV. The start and end points are always given by $k=0$, $k=\infty$. 

For ${\rm Re}\,z<2m_\pi+m_N$ we obtain ${\rm Re}\,z_\sigma <2m_\pi$ and the starting point of the integration at $k=0$ lies below the cut at position (b). Yet, one simply rotates the cut direction further than $-90^0$ and can apply the method as before. 

The analytic continuation of $G_{\sigma N}$ to the lower $z$ half plane, called $G_{\sigma N}^{(2)}$, has been evaluated. A quantitative definition of $G_{\sigma N}^{(2)}$ will be given below [cf. Eq. (\ref{explipath})]. To continue $G_{\sigma N}^{(2)}$ to the upper plane, one can adapt the method of contour deformation, or simply utilize the known continuation in the lower half plane and take advantage of the general analytic property 
\be
G_{\sigma N}^{(2)}(z^*)=[G_{\sigma N}^{(2)}(z)]^*
\label{mirror}
\ee
that holds also for the second sheet. 

\begin{figure}
\includegraphics[width=0.47\textwidth]{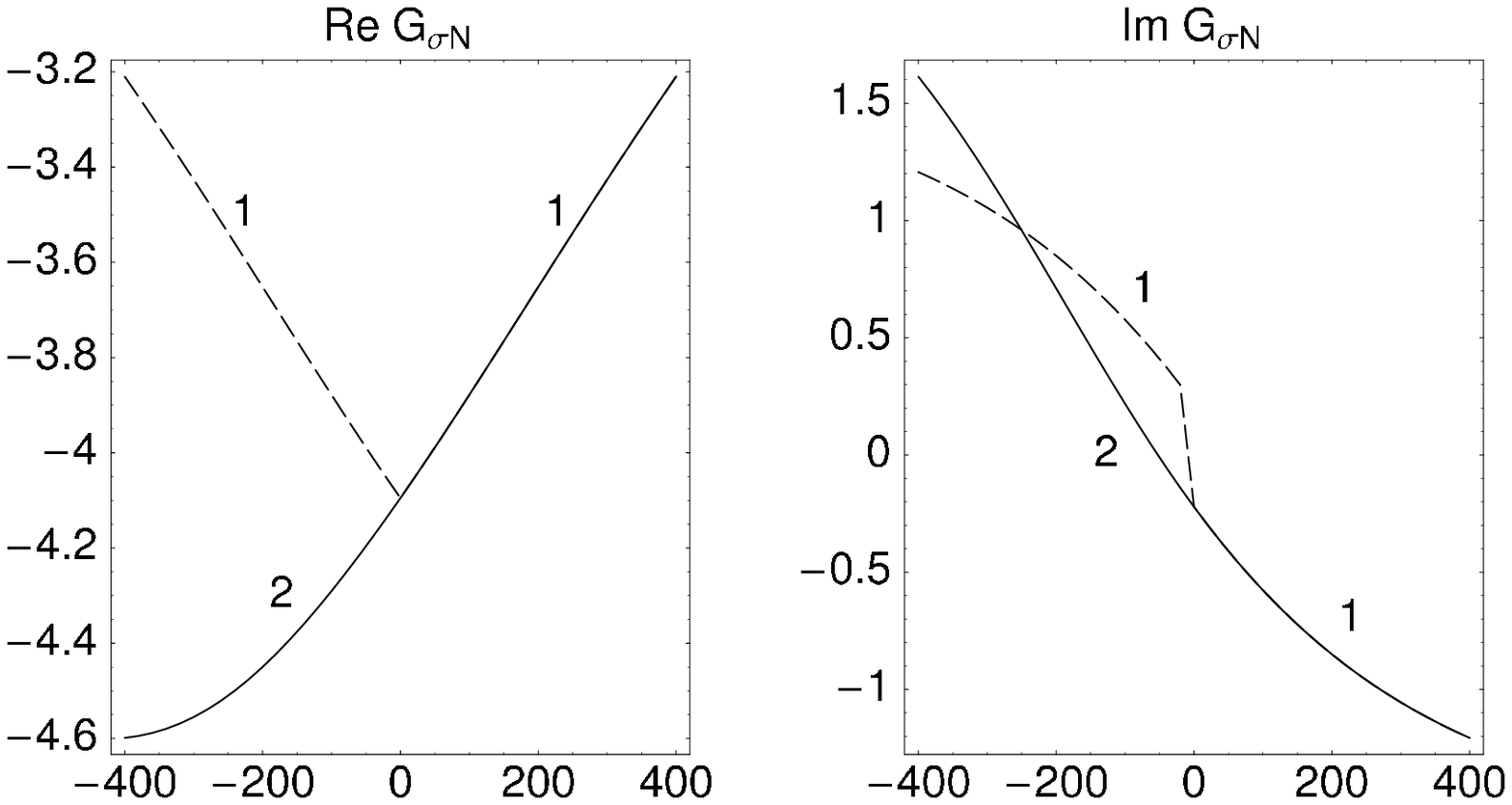}\\

\vspace*{0.3cm}

\hspace{0.1cm}
\includegraphics[width=0.46\textwidth]{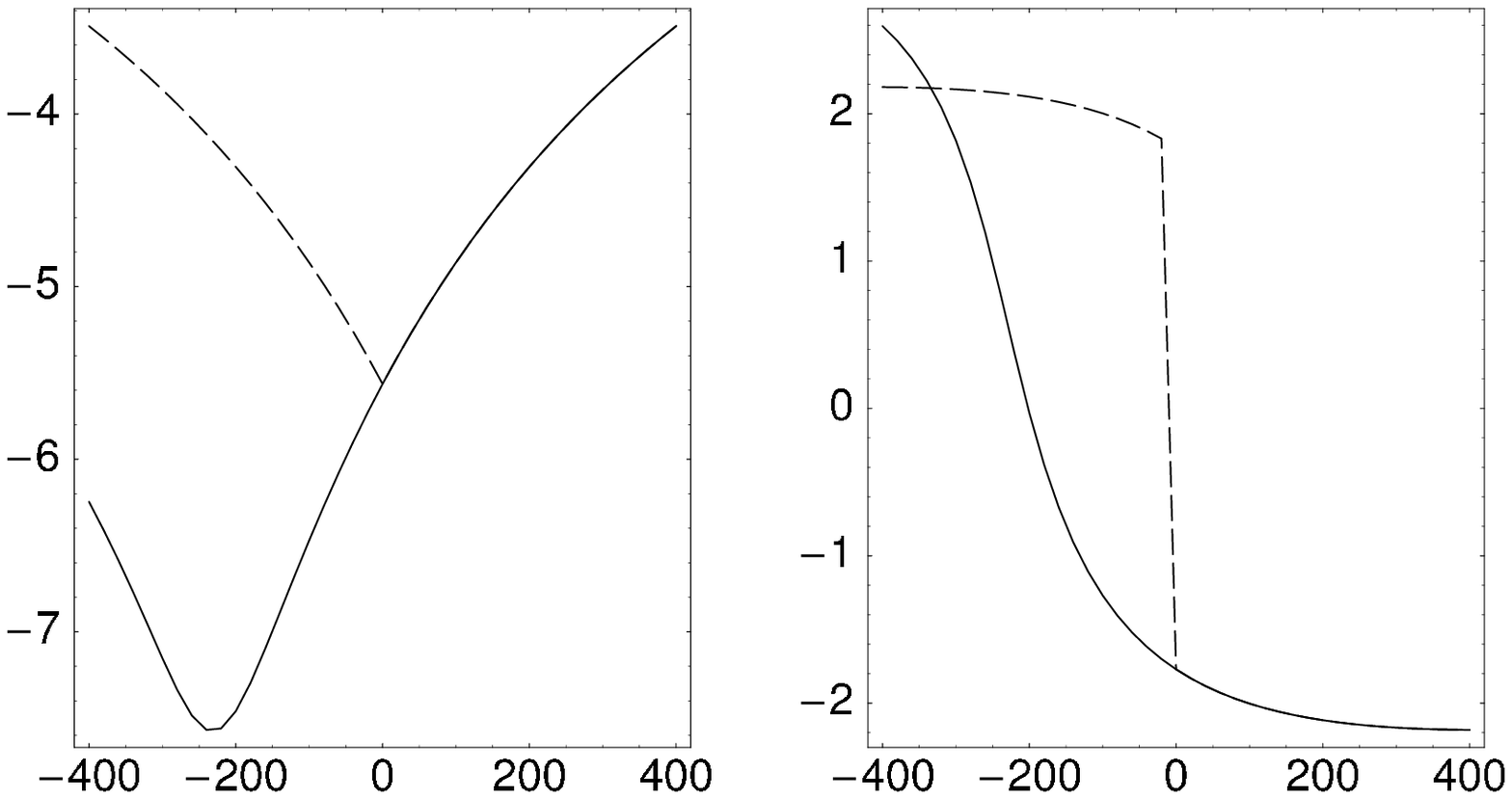}\\

\vspace*{0.1cm}

\includegraphics[width=0.47\textwidth]{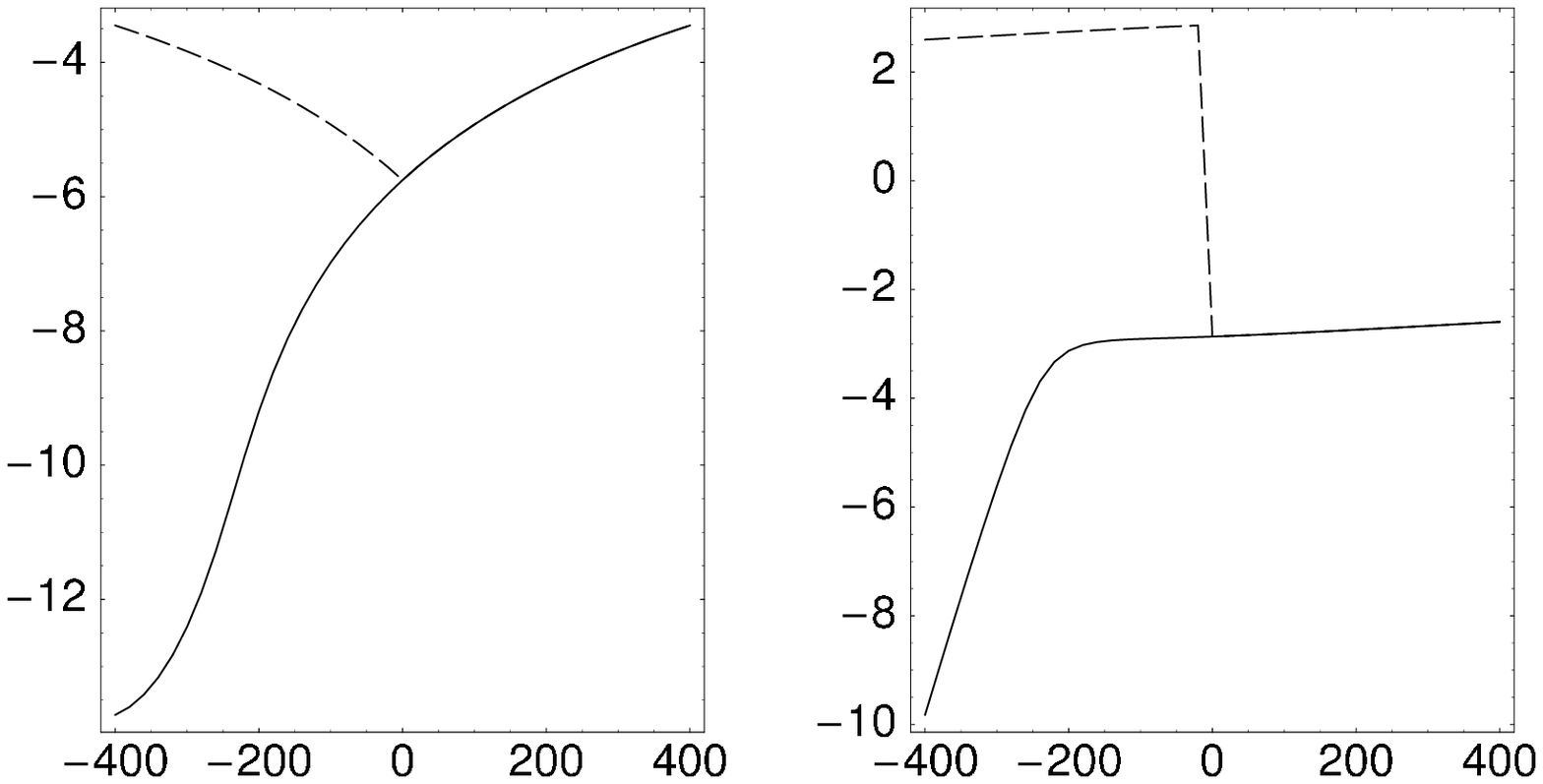}\\

\hspace*{-0.5cm}
\includegraphics[width=0.51\textwidth]{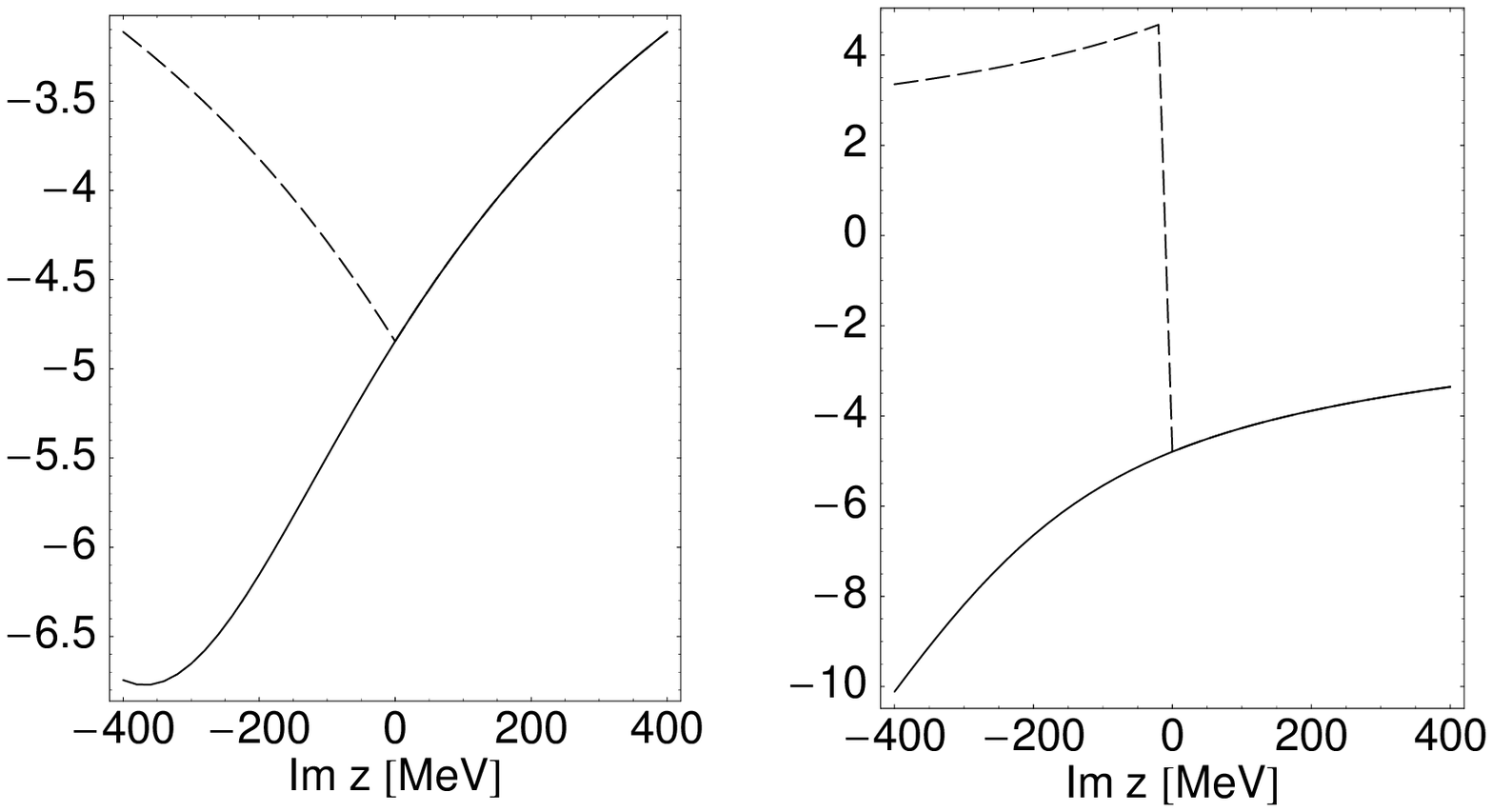}
\caption{Real (left) and imaginary (right) part of $G_{\sigma N}$ [arb. units]. The plots show slices along the imaginary $z$ direction for ${\rm Re}\,z$ fixed at 1.4, 1.75, 1.87, and 2.1 GeV (top to bottom). The solid lines show the analytic continuation to the lower $z$ half plane [$G_{\sigma N}^{(2)}$ for ${\rm Im}\,(z)<0$, $G_{\sigma N}^{(1)}$ for ${\rm Im}\,(z)>0$]. The dashed lines show the first sheet, $G_{\sigma N}^{(1)}\equiv G_{\sigma N}$.}
\label{fig:slices1}
\end{figure}

In Fig. \ref{fig:slices1} we show the analytic continuation as slices along the ${\rm Im}\, z$ direction at fixed ${\rm Re}\, z$.
The solid lines show the first sheet $G_{\sigma N}^{(1)}$ for ${\rm Im}\,z>0$ and $G_{\sigma N}^{(2)}$ for ${\rm Im}\,z<0$. As Fig. \ref{fig:slices1} shows, the transition between both sheets is indeed analytic at ${\rm Im}\,z=0$. The dashed lines show, for ${\rm Im}\,z<0$, the first sheet $G_{\sigma N}^{(1)}$. The cut in $G_{\sigma N}^{(1)}$, situated at ${\rm Im}\,z=0$, appears in Fig. \ref{fig:slices1} as discontinuity for the imaginary part and a non-analyticity (``cusp'') for the real part. The second sheet $G_{\sigma N}^{(2)}$ for ${\rm Im}\,z>0$, obtained through Eq. (\ref{mirror}), is not plotted in Fig. \ref{fig:slices1}.

%%%%%%%%%%%%%%%%%%%%%%%%%%%%%%%%%%%%%%%%%%%%%%%%%%%%%%%%%%%%%%%%%%%%%%%%%%%%%%%%%%%%%%%%%%%%

\subsection{Additional branch points in the complex plane}
\label{sec:foursheets}
The analytic continuation of ${\rm Im}\,G_{\sigma N}$ is shown with the solid lines in the right column of Fig. \ref{fig:slices1}. While, at ${\rm Re}\,z=1.75$ GeV, the imaginary part shows a sharp rise as ${\rm Im}\,z<-250$ MeV, only 120 MeV above at ${\rm Re}\,z=1.87$ GeV the continuation falls rapidly beyond ${\rm Im}\,z< -250$ MeV. The real parts show also rapid changes in this $z$ region.
This is a sign that there is an additional structure. 
In the following we show that the new structure is induced by a branch point. Such additional branch points are known since long \cite{brapothree,Cutkosky:1990zh}.

In Sec. \ref{sec:two-particle} it has been shown for the case of propagators of stable particles that branch points are induced whenever two poles in the complex plane of the momentum integration over $q$ coincide, i.e., the denominator has a double zero in the momentum plane and $q=0$. For the stable case, this situation was given at threshold. In the present case, we have to inspect the denominator of Eq. (\ref{signpro}) and search for poles with respect to $k$. This is, in general, only possible numerically. However, we know that for a branch point, $k=0$. Furthermore, the condition $k=0$ and the fact that we consider the analytic continuation of $G_{\sigma N}^{(2)}$ imply that the $\sigma$ self-energy is evaluated on the second sheet, $\Pi_\sigma^{(2)}$ according to Eq. (\ref{bsheet}) [c.f. Fig. \ref{fig:defo1}]. Then, the denominator of Eq. (\ref{signpro}) has the double zero at
\be
z-m_N-m_\sigma^0-\Pi^{(2)}_\sigma(z-m_N,0)=0.
\ee
On the other hand, the pole of the $\sigma$ itself is given by Eq. (\ref{gsigzwei}); using that $z_\sigma=z-m_N$ at $k=0$, we obtain for the position of the branch point, called $b_2$ in the following,
\be
z_{b_2}=z_p+m_N
\label{polebra}
\ee
where $z_p=875-232\,i$ MeV is the pole position of the $\sigma$ in the complex $z_\sigma$ plane. 

Thus, branch points in the complex $z$ plane of the effective $\pi\pi N$ propagators are directly related to the pole of the unstable particle. An unstable particle $\sigma$ induces branch points in the $\sigma N$ propagator. These branch points are always on the second sheet of $G_{\sigma N}$, as we have seen in the derivation of Eq. (\ref{polebra}), because this is where the resonance poles are. Furthermore, as there is only one pole on the second sheet of the $\sigma$ propagator, there are no further induced branch points of $G_{\sigma N}$ in the $z$ plane, apart from $b_2$ [and $b_2'$ in the upper $z$ half plane due to Eq. (\ref{mirror})].

For the $\rho N$ and $\pi\Delta$ propagators, relation (\ref{polebra}) holds as well, with the corresponding masses and pole positions; in all cases, the validity has been confirmed numerically. 

Relation (\ref{polebra}) has been derived for the specific form of the propagator from Eq. (\ref{signpro}). Yet, the existence of the  branch points $b_2$, $b_2'$ follows in general from the existence of the pole in the scattering amplitude of the unstable particle. Second, the validity of Eq. (\ref{polebra}) does not depend on the special form of the transformation $z_\sigma$ from Eq. (\ref{signpro}): the condition $z_\sigma=z-m_N$ at $k=0$ that led to Eq. (\ref{signpro}) simply reflects the opening of the $\pi\pi N$ threshold at $z=2m_\pi+m_N$ and is obeyed in general~\footnote{The existence of $b_2$ and its general properties are not discussed in Ref. \cite{Suzuki:2008rp}.}.

\begin{figure}
\includegraphics[width=0.45\textwidth]{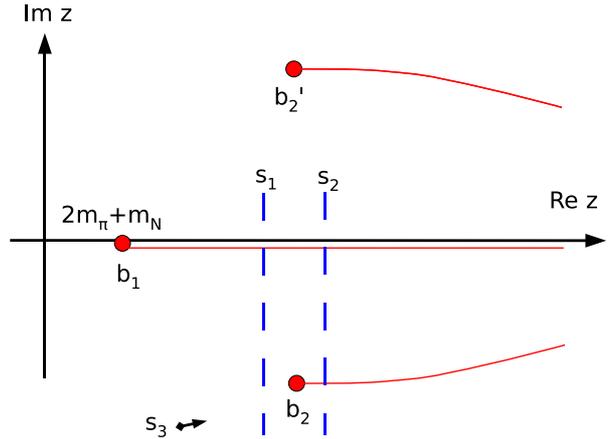}
\caption{The full analytic structure of the $\sigma N$ propagator $G_{\sigma N}$. The branch point $b_1$ is at the $\pi\pi N$ threshold and connects first and second sheet. The branch points $b_2$ and $b_2'$ are located at $z_{b_2}=z_p+m_N$ ($z_{b_2'}=z_p^*+m_N$) and connect second with third, and second with fourth sheet, respectively. The lines $s_1$ and $s_2$ indicate slices plotted in Fig. \ref{fig:allfourslide}. $s_3$ indicates the viewpoint of the plot in Fig. \ref{fig:branchpointb2}.}
\label{fig:sn_structure}
\end{figure}

The resulting analytic structure is shown in Fig. \ref{fig:sn_structure}. 
There is a branch point $b_1$, located at $z_{b_1}=2m_\pi+m_N$, which connects first and second sheet~\footnote{The counting of the sheets refers in this section to the one channel case of $\sigma N$.}, both of them with a cut along the real axis. 
As we have seen previously, this branch point and its cut are induced by the cut of the $\sigma$ self-energy $\Pi_\sigma$.
The additional branch points $b_2$ and $b_2'$ lie in the complex plane, both of them on the second sheet, and induce the two additional sheets three and four. 

\begin{figure}
\includegraphics[width=0.4\textwidth]{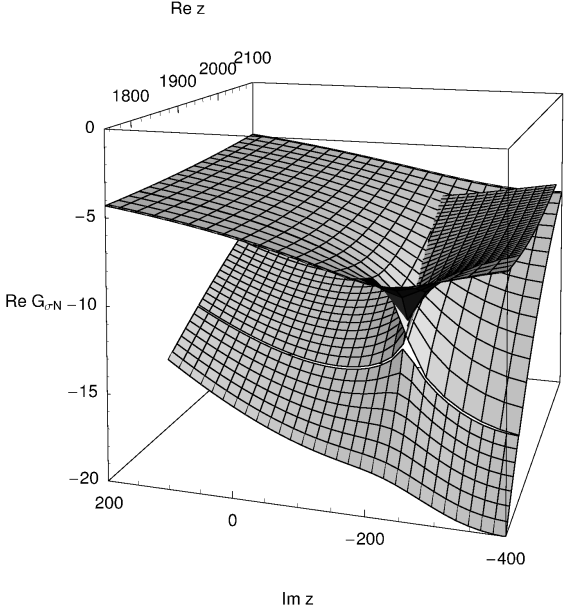}
\caption{Real part of $G_{\sigma N}$ [arb. units] around the branch point $b_2$, as a function of $z$ [MeV]. The intersection of the real parts of the two sheets is visible. The position of this intersection defines the cut plotted in Fig. \ref{fig:sn_structure}.}
\label{fig:branchpointb2}
\end{figure}

In Fig. \ref{fig:branchpointb2} 
we show the real part of $G_{\sigma N}$ around $b_2$ from a viewpoint $s_3$ as indicated in Fig. \ref{fig:sn_structure}. There is an intersection of real parts visible, starting at $b_2$ and extending towards positive ${\rm Re}\, z$ values. For a propagator of stable particles, the real parts of the two sheets intersect along the real axis, or righthand cut, as can be seen in the upper left plot of Fig. \ref{fig:slides_pisig}. It is, thus, straightforward to define the cut belonging to $b_2$ in the same way, i.e. at the intersection of the real parts. This is indicated in Fig. \ref{fig:sn_structure} with the red curved line.

\begin{figure}
\includegraphics[width=0.4\textwidth]{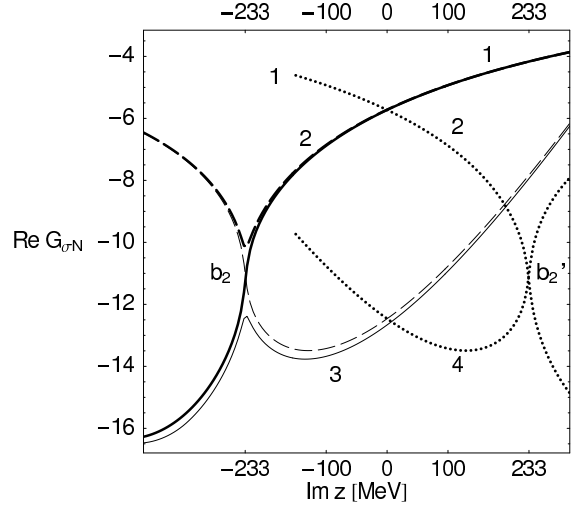}\\

\vspace*{0.5cm}

\includegraphics[width=0.4\textwidth]{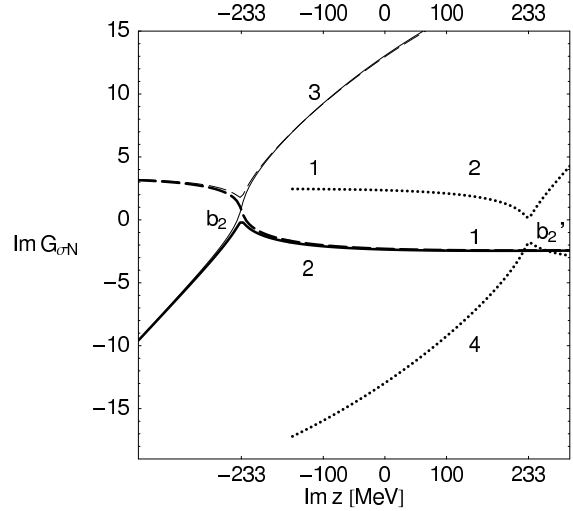}
\caption{The analytic structure of all four Riemann sheets [arb. units], labeled 1 to 4, shown close to ${\rm Re}\, z$ of the branch points $b_2$ and $b_2'$. The thick dashed lines and the thin solid lines show the structures along the slice $s_1$ from Fig. \ref{fig:sn_structure}, i.e. slightly below $b_2$. The thick black solid lines and the thin dashed lines are along $s_2$, i.e., slightly above $b_2$. The dotted lines indicate the presence of a complex conjugate structure around the third branch point $b_2'$.}
\label{fig:allfourslide}
\end{figure}

In Fig. \ref{fig:allfourslide} the four Riemann sheets along the two slices $s_1$ and $s_2$, as indicated in Fig. \ref{fig:sn_structure}, are shown. The slices are located slightly below and above $b_2$. 
In order to understand the structure of the branch point, one can follow paths around it. For example, coming from the real $z$ axis on the second sheet, one can follow $s_1$ and pass by $b_2$ below (thick dashed line in Fig. \ref{fig:allfourslide}). Then, one can move to $s_2$ and move back, towards the real axis, passing $b_2$ above (thin dashed line). As Fig. \ref{fig:allfourslide} shows, one is then not any more on the second sheet, but on sheet 3. Alternatively, one can start at the real $z$ axis on sheet 2, along $s_2$ (thick black solid line), move around $b_2$ and return on $s_1$ (thin solid line), and also get to sheet 3.
Note the presence of a complex conjugate structure that follows from Eq. (\ref{mirror}), associated with the branch point $b_2'$ and indicated with the dotted lines in Fig. \ref{fig:allfourslide}. 

Technically, the transition to the sheets 3 and 4 is achieved by following such paths, while further deforming the integration contour of the $k$ integration. This further deformation is dictated by two requirements: the path in the $k$ plane must not cross the poles induced by the denominator of Eq. (\ref{signpro}), and, second, must not cross the cut along direction (b) from Fig. \ref{fig:defo1} in the $z_\sigma$ plane. 

\begin{figure}
\includegraphics[width=0.39\textwidth]{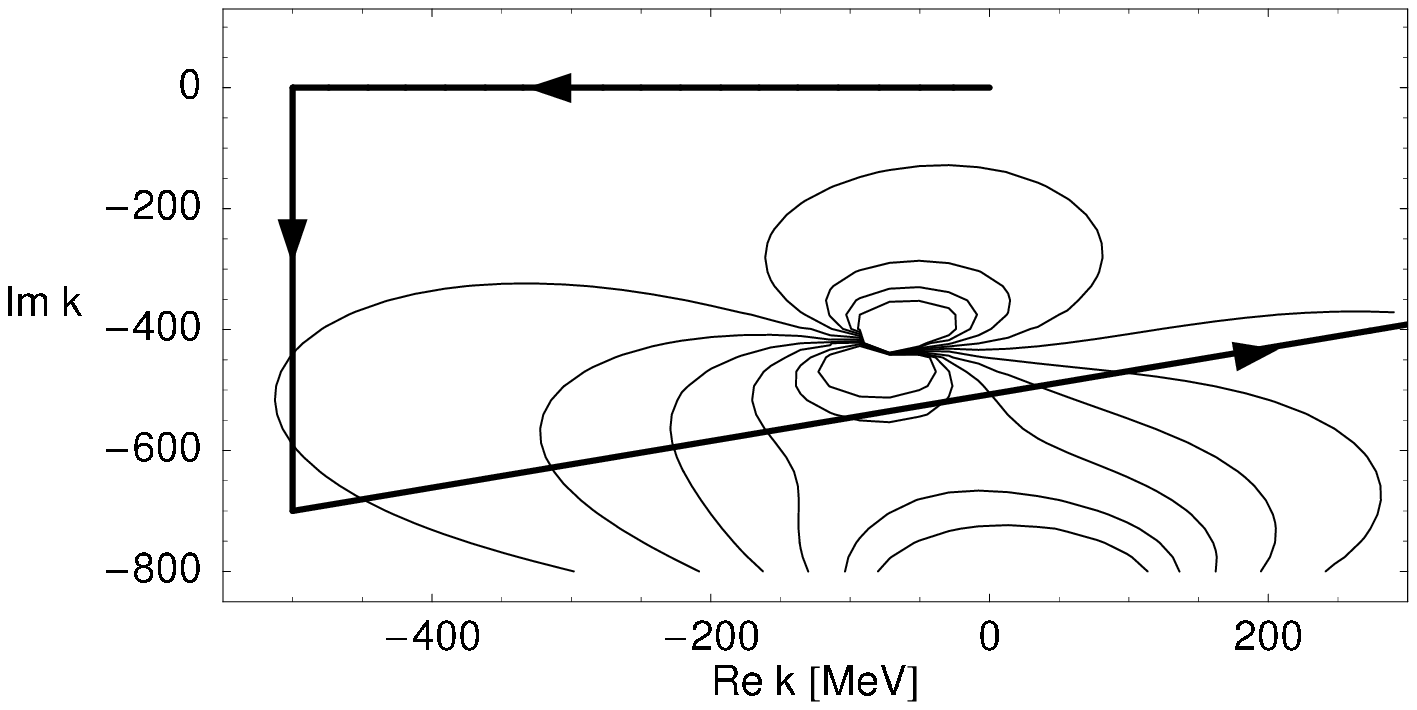}\\

\vspace*{0.3cm}

\includegraphics[width=0.42\textwidth]{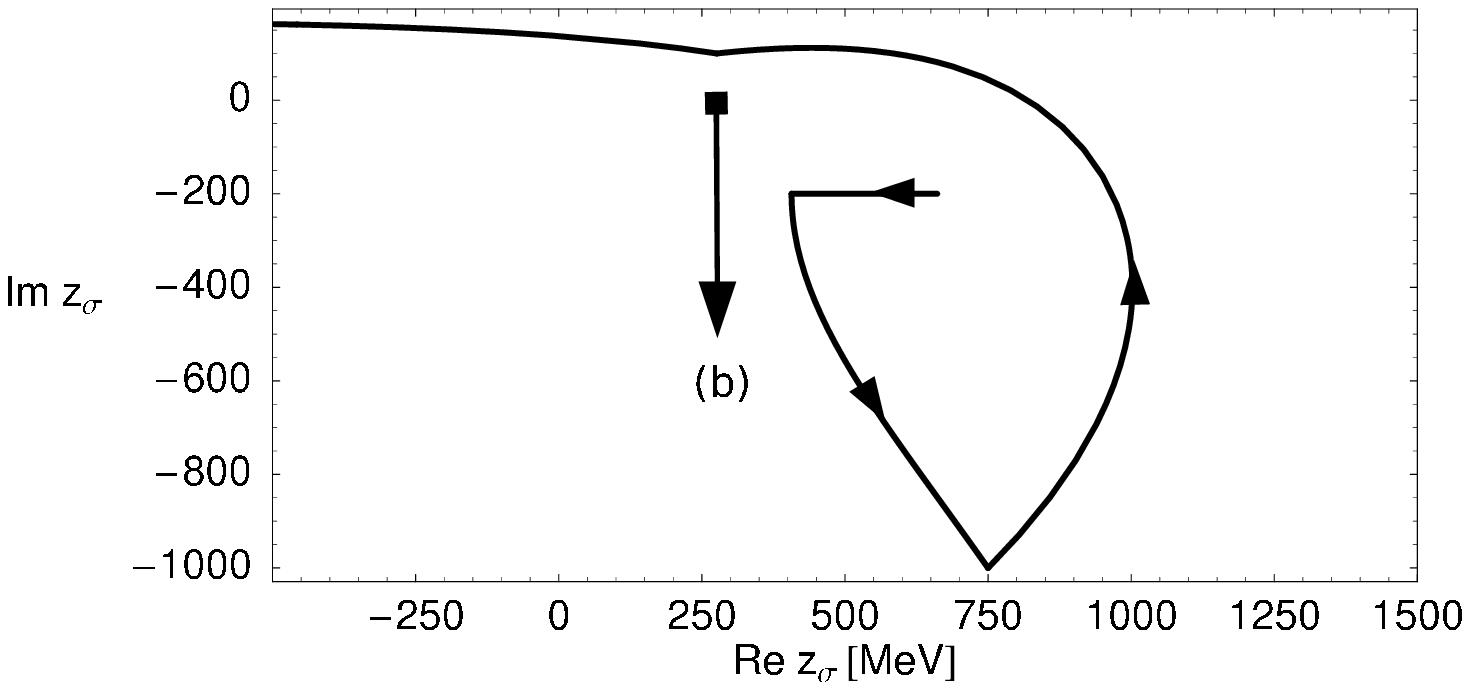}
\caption{Example of a more complicated integration path of the $k$ integration in the $k$ [MeV] and the $z_\sigma$ [MeV] plane. The cut in the $z_\sigma$ plane is in direction (b), same as in Fig. \ref{fig:defo1}. The result in this example  gives the value of $G_{\sigma N}$ at $z=1600-200\,i$ MeV on the third sheet.}
\label{fig:exa_path}
\end{figure}

As an example, we show in Fig. \ref{fig:exa_path} a path in the $k$ plane, together with its image in the $z_\sigma$ plane given by $z_\sigma(z,k)$ from Eq. (\ref{signpro}). With this path, one obtains the continuation to the third sheet along the slice $s_1$ from Fig. \ref{fig:sn_structure}, shown as the thin solid line in Fig. \ref{fig:allfourslide}.
As can be seen in Fig, \ref{fig:exa_path}, in the $k$ plane the path includes the quasi-two-body singularity at $k\sim -70-400\,i$ MeV, while in the $z_\sigma$ plane it does not cross the cut of $\Pi^{(b)}_\sigma$, indicated with the dashed vertical line. Thus, the integrand is always analytic.

It is instructive to discuss the limiting case of a narrow $\sigma$. We define this limit by decreasing the $\sigma\pi\pi$ coupling constant which appears in $v^{\sigma\pi\pi}$ of Eq. (\ref{sigself}). Then, the mass of the $\sigma$ stays approximately constant. As the $\sigma$ becomes narrower, the branch points $b_2$ and $b_2'$ move towards the real axis according to Eq. (\ref{polebra}). Simultaneously, the discontinuity on the real axis, associated with $b_1$, becomes weaker. In the limit of zero width, $b_2$ and $b_2'$ coincide on the real axis at $z=m_N+m_\sigma^0$, and the two associated cuts coincide as well and run along the real axis. The branch point $b_1$ and its cut vanish, i.e., sheet 1 and 2 coincide below $z=m_\sigma^0+m_N$. 

Also, the other sheets overlay: We consider the real parts of $G_{\sigma N}$ in Fig. \ref{fig:allfourslide}. They form approximately two ``x'' with the centers at $b_2$ and $b_2'$. As the $\sigma$ width approaches zero, the ``x'' become symmetric and for zero width overlap exactly, reducing the number of Riemann sheets from four to two; the real part has then the same structure as in the second row, left, of Fig. \ref{fig:slides_pisig}. Thus, in the limit of zero $\sigma$ width, we have precisely the sheet structure of the propagator of stable particles as shown in Figs. \ref{fig:sisheets} and \ref{fig:slides_pisig}, with one branch point at $z=m_\sigma^0+m_N$, one righthand cut, and two sheets.

In Fig. \ref{fig:allfourslide} we show the labeling of the Riemann sheets. The distinction of sheet 1 and 2 is clear: While the physical sheet 1 is obtained with an integration along an undeformed path in Eq. (\ref{signpro}), sheet 2 is the analytic continuation of sheet 1 along the righthand $\pi\pi N$ cut. The distinction between sheet 2 and 3 is defined as follows. We have just argued that a natural choice of the cut associated with $b_2$ is along the intersection of the real parts (cf. Fig. \ref{fig:branchpointb2}). This also helps us understand the limiting case of a narrow $\sigma$. Yet, other choices are possible. For practical reasons, we prefer a cut that delivers analytic slices along the ${\rm Im}\, z$ direction, such as plotted in Fig. \ref{fig:slices1}. For this, we have to put the cut of $b_2$ into the negative ${\rm Im}\, z$ direction. This is the definition which we will adopt for sheet 2: 

Sheet 2, in the lower half plane, is the sheet that can be reached along straight paths into the negative ${\rm Im}\, z$ direction, starting from the real $z$ axis at the continuation of sheet 1. The additional Riemann surface induced by $b_2$, not reachable by such paths, is then sheet 3. Sheet 2 in the upper half plane, together with the branch point $b_2'$, can be obtained from sheet 2 in the lower half plane using Eq. (\ref{mirror}). Accordingly, the cut associated with $b_2'$ is then in the positive ${\rm Im}\, z$ direction. Sheet 4, associated with the branch point $b_2'$, is defined accordingly and can be obtained from sheet 3 using Eq. (\ref{mirror}).

%%%%%%%%%%%%%%%%%%%%%%%%%%%%%%%%%%%%%%%%%%%%%%%%%%%%%%%%%%%%

\subsection{Formalism of path deformation}
\label{sec:formalpath}
With these definitions, the paths which lead to the various sheets are formally defined. For simplicity, all paths are chosen piecewise linear in the $k$ plane. We write for sheet $(j)$:
\be
&&G_{\sigma N}^{(j)}(z)=\int_{\Gamma^{(j)}} dk \,k^2\, F(k)\, g_{\sigma N}^{(j)}(z,k)\non
&=&\sum_{i=1}^{n_j}\int\limits_0^1 dt\,(k_{i}-k_{i-1})\,[k_i(t)]^2\,F(k_i(t))\,g_{\sigma N}^{(j)}(z,k_i(t))\non
\label{explipath}
\ee
with $g_{\sigma N}$ from Eq. (\ref{signpro}). Here, $g_{\sigma N}^{(1)}$ [$g_{\sigma N}^{(2,3)}$] is evaluated with $\Pi_\sigma$ from Eq. (\ref{sigself}) [$\Pi_\sigma^{(b)}$ from Eq. (\ref{bsheet})]. In Eq. (\ref{explipath}), $k_i(t)=k_{i-1}+(k_{i}-k_{i-1})\,t$. The $k_i$ are the edge points of the paths and shown in Tab. \ref{tab:paths} for each Riemann sheet $(j)$. For the physical sheet $j=1$, $k_0=0$ and $k_{n_j=1}\to\infty$, i.e. the integration is along a straight path from zero to infinity in the $k$ plane. For the sheets $j=2,3$, the first, before last, and last point in the $k$ plane are given by
\be
k_0			&=&0, \non
z_{\rm eff}(k_{n_j-1},z)	&=& 
\begin{cases}
2m_\pi+100\,i\,\text{MeV} & \text{for $\sigma N$}\\
2m_\pi+200\,i\,\text{MeV} & \text{for $\rho N$}\\
m_\pi+m_N+100\,i\,\text{MeV}& \text{for $\pi \Delta$}
\end{cases}\non
k_{n_j}			&\to &\infty.
\label{startend}
\ee
where $z_{\rm eff}$ is the energy of the unstable particle $\sigma,\rho$, or $\Delta$, given by Eq. (\ref{signpro}) for the $\sigma$ case. For the corresponding expressions $g_{\rho N}$, $g_{\pi\Delta}$ as they appear in Eq. (\ref{explipath}) for the $\rho N$ and $\pi\Delta$ propagator, and for the transformations $z_\rho$ and $z_\Delta$, see Refs. \cite{Schutz:1998jx,Krehl:1999km}.

\begin{figure}
\includegraphics[width=0.48\textwidth]{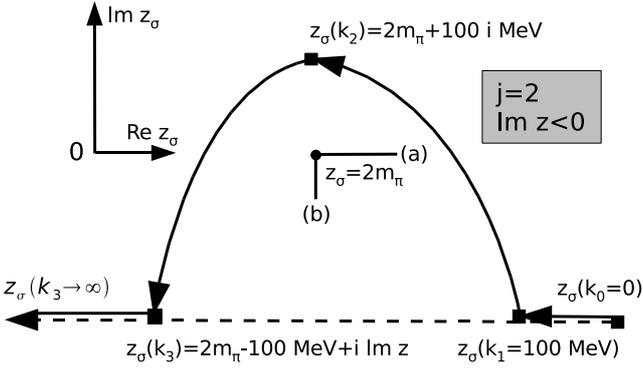}
\caption{Integration contour to calculate $\delta G$. The plot shows the image $z_\sigma(k,z)$ in the $z_\sigma$ plane. See text for further explanations.}
\label{fig:implement}
\end{figure}

\begin{table}
\caption{Parameterization of the paths in the $k$-plane for sheets $j=2,3$, according to Eq. (\ref{explipath}). The list shows the positions [MeV] of intermediate points in the $k$ plane. For the other points $k_0$, $k_{n_j-1}$, $k_{n_j}$, see Eq. (\ref{startend}). Sheet 4 is obtained from sheet 3 through Eq. (\ref{mirror}).}
\begin{tabular}{llll}
 \hline\hline
\hspace*{1.7cm}&$\sigma N$\hspace*{1.8cm}&$\rho N$\hspace*{1.8cm}&$\pi\Delta$	\\
\multicolumn{3}{l}{$j=2$, ${\rm Re}\,z<{\rm Re}\,z_{b_2}$}& 		\\
$k_1$			&$100$ 		&$100$ 		&--- 		\\
\multicolumn{3}{l}{$j=2$, ${\rm Re}\,z\geq{\rm Re}\,z_{b_2}$}& 		\\
$k_1$			&$1-300\,i$ 	&$1-500\,i$ 	&$1-300\,i$ 	\\
\multicolumn{3}{l}{$j=3$, ${\rm Re}\,z<{\rm Re}\,z_{b_2}$}& 		\\
$k_1$			&$-500$	 	&$-500$	 	&$-80-i$ 	\\
$k_2$			&$-500-700\,i$  &$-500-700\,i$  &$-80-200\,i$   \\
\multicolumn{3}{l}{$j=3$, ${\rm Re}\,z\geq {\rm Re}\,z_{b_2}$}& 		\\
$k_1$			&$1+500\,i$	&$1+500\,i$	&$30+80\,i$	\\
$k_2$			&$500+500\,i$   &$500+500\,i$   &$100+70\,i$	\\
\hline\hline
\end{tabular}
\label{tab:paths}
\end{table}
The distinctions ${\rm Re}\,z\lessgtr {\rm Re}\,z_{b_2}$ in Tab. \ref{tab:paths} ensure that the cut of $b_2$ is along the negative ${\rm Im}\,z$ direction, as discussed at the end of the previous section. As an example, we show in Fig. \ref{fig:implement}
with the solid line the path for the second sheet of the $\sigma N$ propagator, case ${\rm Re}\,z < {\rm Re}\,z_{b_2}$ in the $z_\sigma$ plane. Comparing this figure with Fig. \ref{fig:defo1}, an additional structure induced by $z_\sigma(k_1=100\,\text{MeV})$ is visible; this additional edge point is necessary to avoid mixing of sheet 2 and 3. For the endpoint $z_\sigma(k_3)$ that is finite in Fig. \ref{fig:implement}, rather than $z_\sigma(k_3\to\infty)$ as Eq. (\ref{startend}) prescribes, see the discussion on implementation in the J\"ulich model in Sec. \ref{sec:implement}.

The paths for the sheets $j=2,3$ lead to the correct sheets, but only in a certain range for the total energy $z$. Sheet (2), with the above defined paths, is calculated correctly for $2m_\pi+m_N<{\rm Re}\,z<2500$ MeV and $-400<{\rm Im}\,z<0$ MeV, for $\sigma N$, $\rho N$, and $\pi\Delta$. For the sheets (3), the continuations are valid around the branch points $b_2$; for $\sigma N$,  the range of applicability has been tested to be $1600<{\rm Re}\,z<2100$ MeV and $-400<{\rm Im}\,z<+300$ MeV. For $\rho N$, $1500<{\rm Re}\,z<2000$ MeV and $-400<{\rm Im}\,z<0$ MeV; for $\pi\Delta$, $1250<{\rm Re}\,z<1450$ MeV and $-250<{\rm Im}\,z<0$ MeV. If one wishes to obtain the sheets (3) outside these areas, one may have to redefine the paths; the pole in the $k$ plane moves as a function of $z$ and for $z$ outside these areas, it may cross the integration contour.

The $\rho N$ and $\pi\Delta$ propagators have the same analytic properties as the $\sigma N$ propagator. In particular, the poles of the $\rho$ and the $\Delta$ induce additional branch points $b_2,\,b_2'$ of $G_{\rho N}$ and $G_{\pi\Delta}$ in the complex $z$ plane. For these channels, these branch points are even more relevant for the analytic continuation because the $\rho$ and the $\Delta$ are narrower than the $\sigma$ and the branch points lie close to the real axis according to Eq. (\ref{polebra}). The branch points $b_2$ and $b_2'$ have large numerical effects in their surroundings as can be seen in Figs. \ref{fig:slices1} to \ref{fig:allfourslide} and, thus, play an important role for the analytic continuation. The integration paths to obtain the different sheets of the $\rho N$ and $\pi\Delta$ propagators are given in Table \ref{tab:paths}, together with the points defined in Eq. (\ref{startend}).

To finish this discussion, let us again point out the importance to control the cuts induced by the branch points $b_2,\,b_2'$. As discussed above, the condition ${\rm Re}\,z\lessgtr {\rm Re}\,z_{b_2}$ controls the cut. In particular, in these two different cases the $k$ integration contour passes by the quasi-two-particle singularity above and below in the $k$ plane, respectively. This distinction is mandatory; if the contour passes for all $z$ above or for all $z$ below that singularity one evaluates the third sheet instead of the second for some $z$; poles on that sheet are much less relevant than poles on the second sheet, because the third sheet is not directly connected to the physical axis.

%%%%%%%%%%%%%%%%%%%%%%%%%%%%%%%%%%%%%%%%%%%%%%%%%%%%%%%%%%%%%%%%%%%%%%%%%

\subsection{Implementation of the continued $\sigma N$, $\rho N$, $\pi\Delta$ sheets}
\label{sec:implement}
To obtain the amplitude of the J\"ulich model on the different $\sigma N$, $\rho N$, $\pi\Delta$ sheets, we add the difference between second and first sheet, similar as in the case of the channels $\pi N$ and $\eta N$ in Eq. (\ref{analpin}). In Sec. \ref{sec:two-particle} we have seen that for the $\pi N,\,\eta N$ propagators the prescription to obtain the second sheet consists in adding the discontinuity twice. In the present case of the unstable $\sigma N, \,\rho N,\,\pi\Delta$ propagators, one can proceed similarly. Yet, for the effective $\pi\pi N$ channels, this implies an approximation which will be discussed below.

In Fig. \ref{fig:implement} the image $z_\sigma(k,z)$ of the $k$-integration in the $z_\sigma$ plane is shown. Consider the case $j=2$ for the $\sigma N$ propagator, i.e. the second sheet. The dashed line shows the integration contour for $j=1$, the first sheet. In the $k$ plane, this integration is along a straight path from $k=0$ to $k=\infty$. 

The solid lines show the integration according to the case $j=2$ with the edge points $k_0,\,k_1$, and $k_2$ given in Eq. (\ref{startend}) and Table \ref{tab:paths}. 
For the endpoint $k_3$, we have chosen, instead of $k_3\to\infty$, an intermediate point $z_\sigma(k_3)$ that is on the contour of the integration for the first sheet and below the $\pi\pi$ threshold (remember that for the second sheet, the $\pi\pi$ cut is in direction (b) of Fig. \ref{fig:implement}). In the difference of second and first sheet, the remaining path from the point $k_3$ to $\infty$ cancels. 

For the plots of numerical results in this section, we introduced a form factor $F$ in Eq. (\ref{signpro}) to regularize the integral. For the implementation in the J\"ulich model, this artificial form factor is removed; the result for the difference of sheets is finite anyways as discussed before. 

Thus, the difference between sheet $(j)$ and sheet (1) is given by 
\be
\delta \tilde{G}_{\rm eff}^{(j)} &=&(G_{\rm eff}^{(j)}-G_{\rm eff}^{(1)})_{F=1},
\label{deltag}
\ee
where $j=2,\,3$. Here and in the following, the subscript ``eff'' indicates the effective $\pi\pi N$ channels. The end points of both integrations for $G_{\rm eff}^{(j)}$ and $G_{\rm eff}^{(1)}$ are at
\be
&&z_{\rm eff}(k_{n_j},z)	\non 
&=&\begin{cases}
2m_\pi-100\,\text{MeV}+i\,{\rm Im}\,z & \text{($\sigma N$, $\rho N$)}\\
m_\pi+m_N-100\,\text{MeV}+i\,{\rm Im}\,z& \text{($\pi \Delta$)}
\end{cases}
\ee
which replaces $k_{n_j}$ from Eq. (\ref{startend}). As Fig. \ref{fig:implement} shows, the overall integration path to obtain $\delta \tilde{G}$ is a closed contour, $\delta G=\oint dk\,f(k,z)$, but with start and end point on different sheets of the self-energy of the unstable particle.

The next step to implement the sheets of the effective $\pi\pi N$ propagators in the J\"ulich model is similar to Eq. (\ref{analpin}),
\begin{multline}
\langle q_{cd}|T^{(j)}-V|q_{ab}\rangle=\\
\delta G_{\rm eff}^{(j)}+\int dq_\uns\,q_\uns^2\frac{\langle q_{cd}|V|q_\uns\rangle\langle q_\uns|T^{(j)}|q_{ab}\rangle}{z-E_\uns-\Pi_\uns}\nonumber
\end{multline}
\be
\delta G_{\rm eff}^{(j)}=\delta \tilde{G}_{\rm eff}^{(j)}\,\langle q_{cd}|V|q_{\rm{on}}^>(\uns)\rangle\langle q_{\rm{on}}^>(\uns)|T^{(j)}|q_{ab}\rangle.
\label{analun}
\ee
As in case of the channels with stable particles, $\delta G$ is added at the on-shell point. In case of the channels with stable particles, this is exact because the $\delta$ function that evaluates the imaginary part puts the vertices automatically on-shell [cf. Eq. (\ref{prescristable})]. 

In case of the effective $\pi\pi N$ channels, this factorization of the last line in Eq. (\ref{analun}) is an approximation. For $q_{\rm{on}}^>(\uns)$, we choose the on-shell momenta of the kinematics
\begin{align}
q_{\rm{on}}^>(\uns)&=q_{\rm{on}}^>(z\to m_N+ n\,m_\pi)&\quad (\sigma N, \,\rho N)\non
q_{\rm{on}}^>(\uns)&=q_{\rm{on}}^>(z\to m_\pi+ m_\Delta)&\quad (\pi\Delta)
\end{align}
with $m_\Delta=1232$ MeV and $m_N,\,m_\pi$ the nucleon and pion mass, respectively. We test the cases $n=2,\,3$ for all results. The pole positions and residues are very similar in both cases. 

Another motivation for the factorization is given by the following considerations. 
The kinematic $z\to m_N+ 2\,m_\pi$  for the $\sigma N$ (and also $\rho N$) propagator corresponds in good approximation to the maximum of ${\rm Im}\,k^2\,g_{\sigma N}$ from Eq. (\ref{signpro}), which gives the discontinuity of the $\pi\pi N$ cut along the physical axis. Strictly speaking, the discontinuity receives contributions from ${\rm Im}\,k^2\,g_{\sigma N}$ for all $k$, but the distribution is still concentrated around the maximum due to the factors $k^2$ and the form factors, that suppress the contributions for small and large $k$ values, respectively.

Furthermore, the kinematics $z\to m_N+ 2\,m_\pi$ or $z\to m_N+ 3\,m_\pi$ corresponds to typical $2\pi$ invariant masses in $\pi N\to\pi\pi N$ in the second resonance region in $\pi N\to\pi\pi N$, where the branching ratios into $\rho N$ and $\sigma N$ are analyzed.

The principal difficulty in going beyond the factorization of $\delta \tilde{G}_{\rm eff}^{(j)}$ in Eq. (\ref{analun}) is the incompatibility of the integration paths that lead to unphysical sheets of the effective $\pi\pi N$ propagators, and the position of three-body cuts in the transition potentials $V$ that require a different integration path. While the access to the different sheets is dictated by the deformed paths as discussed above, the momentum integration that respects three-body cuts is from $k=0$ along a straight path rotated into the complex plane~\cite{Gasparyan:2003fp}. A unified treatment is beyond the scope of this work, although possible in principle.

The analytic continuation of $T$ for the first sheet, as given by the integral terms of Eqs. (\ref{analpin}) and (\ref{analun}), is restricted by the two-body cuts of the stable $\pi N$ and $\eta N$ propagator, as well as the pseudo-two-body cuts of $\sigma N$, $\rho N$ and $\pi\Delta$. These zeros of the propagator denominators $z-E_1-E_2$, together with the rotation of the integration path into the lower complex $q$ half plane~\cite{Gasparyan:2003fp}, induce fallacious non-analyticities in the lower $z$ half plane. Thus, we restrict the analysis of the analytic continuation to the upper $z$ half plane. Yet, for all results of this study in Sec. \ref{sec:results}, we will quote the values for pole positions, residues etc. in the lower $z$ half plane that are easily obtained using Eq. (\ref{mirror}).

Additionally, the three-body cuts of the potentials $V$ (in particular, the pion exchange), together with the rotated integration path, induce similar structures for the first sheet in  the upper $z$ half plane. Since these structures are located at ${\rm Im}\,z>150,\,200$ MeV for all $z$ we can safely search for resonances up to a width of $\Gamma=300$ or 400 MeV, and this issue is of no relevance here. For energies above the second resonance region, even larger values for ${\rm Im}\,z$ are accessible. Within our formalism, an exchange propagator is of the form $1/[E_x(z-E_x-E_1-E_2+i\epsilon)]$. There are kinematic factors $1/E_x$ of the exchanged particle $x$. They induce similar structures as the three-body cuts but they are also situated at ${\rm Im}\,z>150,\,200$ MeV.
We make no attempt to access the sheets induced by the three-body cuts associated with the exchange potentials $V$. These sheets are far away from the physical axis and structures on those sheets can only have minor impact on physical scattering. For completeness, the short nucleon, circular and other cuts, associated with nucleon exchange and crossing symmetry, are discussed in Appendix \ref{sec:app1}. It is shown that this cut structure is indeed present in the model and can be identified with different ingredients of the model.

%%%%%%%%%%%%%%%%%%%%%%%%%%%%%%%%%%%%%%%%%%%%%%%%%%%%%%%%%%%%
%%%%%%%%%%%%%%%%%%%%%%%%%%%%%%%%%%%%%%%%%%%%%%%%%%%%%%%%%%%%

\section{Results}
\label{sec:results}
\subsection{Selection of Riemann sheets}
\label{sec:selectsheets}
In Sec. \ref{sec:anal} the analytic structures of the propagators of channels with stable particles $\pi N$, $\eta N$ and of effective $\pi\pi N$ channels $\pi\Delta$, $\rho N$, and $\sigma N$ have been determined. The analytic continuations of the propagators determine the analytic continuation of the $T$ matrix. In this section, we determine the properties of $T(z)$ in terms of poles and zeros in the complex $z$ plane.

For a channel with stable particles, there are two sheets, while for unstable particles, there are four as we have seen in the previous section. Thus, for the channel space considered here, there are $2^2 4^3=256$ sheets corresponding to the two stable and three effective $\pi\pi N$ channels.

Poles can be located on all sheets. Yet, depending on the sheet, their influence at the real $z$ axis (``physical axis'') is different: For a channel of stable particles with two sheets, consider a virtual state. This is a pole at $z=z_0$ on the second (``unphysical'') sheet in the lower $z$ half plane, and below threshold, i.e. ${\rm Re}\,z_0<z_{\rm thres}$. Then, the closest point to the physical axis is $z_{\rm thres}$ [c.f. Fig. \ref{fig:slides_pisig}, third row]. Because the pole contribution has a $1/(z-z_0)$ energy dependence, such a pole can only appear as a structure at $z_{\rm thres}$, e.g. as a threshold enhancement. It cannot create a resonance shape on the physical axis at $z={\rm Re}\,z_0$. 

Thus, for the pole search with respect to the $\pi N$ and $\eta N$ channels, below threshold one chooses the first sheet of the propagator $G$. This means looking for bound states with respect to that channel. Above threshold, one chooses the second sheet. Formally, this can be written as
\be
G&=&
\begin{cases}
G^{(2)}	&	\text{if Re $z\geq z_{\rm thres}$}\\
G^{(1)}	&	\text{if Re $z<z_{\rm thres}$}
\end{cases}\non
\label{defstableg}
\ee
where $z_{\rm thres}=m+M$ is the threshold energy. This choice will induce cuts in the amplitude along the ${\rm Im}\,z_0$ direction, for ${\rm Re}\,z=m_\pi+m_N$ and ${\rm Re}\,z=m_\eta+m_N$. 

As pointed out in Sec. \ref{sec:foursheets}, for each effective $\pi\pi N$ channel $\pi\Delta$, $\rho N$, and $\sigma N$ there are four sheets. For the pole search, we choose the unphysical sheet with respect to the $\pi\pi N$ cut, i.e., the second sheet according to the definitions at the end of Sec. \ref{sec:foursheets}. In particular, the second sheet is reachable from the physical axis via straight paths in the negative ${\rm Im}\, z$ direction. There are additional sheets induced by branch points $b_2,\,b_2'$ in the complex plane as we have seen in Sec. \ref{sec:foursheets}. Yet, coming from the physical axis, those sheets are only accessible via paths around these branch points and poles on them have little influence on the physical axis [cf. Fig. \ref{fig:sn_structure}]. This is in analogy to the case of the $\pi N$ and $\eta N$ propagators discussed previously. 

\begin{figure}
\includegraphics[width=0.48\textwidth]{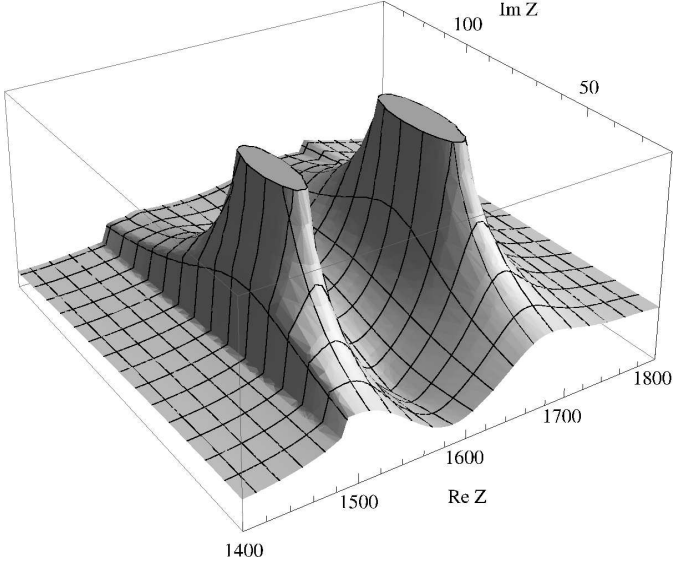}
\caption{Modulus $|T(S_{11})|$ [arbitrary units] as a function of the complex scattering energy $z$ [MeV]. The poles of the $N^*(1535)$ and $N^*(1650)$ are clearly visible. Also, one sees a discontinuity along ${\rm Re}\,z=m_\eta+m_N$, coming from the prescription of Eq. (\ref{defstableg}) for the $\eta N$ channel. Behind the $N^*(1650)$, one sees a cut lying in the positive ${\rm Im}\, z$ direction, induced by the branch point $z_{b_2'}=m_N+z^*_\rho$ of the $\rho N$ channel.}
\label{fig:s11_3d}
\end{figure}

The selection criteria discussed above define one out of 256 sheets, where poles are searched. We refer to this as ``second sheet'' of $T$ in the following. As an example of the analytic structure, Fig. \ref{fig:s11_3d} shows $|T|$ on that sheet, in the $S_{11}$ partial wave in $\pi N\to\pi N$.
The two poles associated with the $N^*(1535)$ and the $N^*(1650)$ are clearly visible. Yet, there are additional structures: First, a discontinuity along ${\rm Re}\,z=m_\eta+m_N$ coming from the prescription of Eq. (\ref{defstableg}). Second, one sees a similar discontinuity starting behind the $N^*(1650)$ pole resulting from the definition of the second $\rho N$ sheet: the branch cut is in the positive ${\rm Im}\, z$ direction and starts at the branch point $z_{b_2'}=z_\rho^*+m_N\sim 1700+64\,i$ MeV where $z_\rho=763-64\,i$ MeV is the pole position of the $\rho$ [cf. Eq. (\ref{polebra})]. 

The first sheet is free of poles as we have checked; we search for poles on the second sheet. Yet, there are poles on other sheets which we comment on in the following. Consider a pole that couples only weakly to a given channel, i.e. its residue to e.g. the $\rho N$ channel is small. Suppose the pole has been found on the second sheet. Then, at the pole position, the term $(1-VG)^{-1}$ which appears in the solution of Eq. (\ref{bse}) is singular. We consider an element of this matrix and write symbolically, with $G_{\rho N}^{(2)}$ the $\rho N$ propagator on the second sheet,
\be
(1-VG)=a+ b\, G_{\rho N}^{(2)}=0
\ee
omitting further indices, sums and integrations.
$a$ contains the terms with intermediate states of other channels. The weak coupling to $\rho N$ is reflected by the fact that the $b$ terms are small compared to the $a$ terms, and the replacement $G_{\rho N}^{(2)}\to G_{\rho N}^{(3)}$ does not change much the position of the zero; the resonance pole will reappear on the third or even fourth $\rho N$ sheet. 

Such replica of poles on other sheets have no physical implications. E.g., the $\Delta^*(1700)$ has a pole at $z_0=1637-118\,i$ MeV on the second $\rho N$ sheet and a replica on the third one, just a few MeV away from $z_0$ and with almost the same residues. Even the distance from both pole positions to the physical axis via paths over analytic regions of the amplitude is approximately the same. Yet, one of the poles is sufficient to describe the $\Delta^*$ properties on the physical axis to high accuracy~\cite{Doring:2009bi}. The two-pole structure is, in this case, trivial. 

The situation is different for e.g. the proposed two-pole structure of the $\Lambda(1405)$~\cite{Jido:2003cb}. In the latter case, the two poles lie on the same sheet on different positions and have quite different residues to the different channels. While there is evidence \cite{Magas:2005vu} for the two-pole structure of the  $\Lambda(1405)$, the trivial replica found here have no physical consequences.

Yet, in the present context, we sometimes find a pole on the third sheet of $\rho N$, but no counterpart on the second sheet; in that case the coupling of the state to $\rho N$ is strong. In such a situation, discussed in Sec. \ref{sec:pol3rhon}, a pole on the third $\rho N$ sheet can indeed induce visible structures on the physical axis that cannot be explained from the amplitude on the second sheet.

%%%%%%%%%%%%%%%%%%%%%%%%%%%%%%%%%%%%%%%%%%%%%%%%%%%%%%%%%%%

\subsection{Pole positions and residues}
\label{sec:respa}
The scattering amplitude $\tau$ for the transition $i\to f$ in channel space is connected to the amplitude $T$ of Eq. (\ref{bse}) by
\be
\tau_{fi}&=&-\pi\sqrt{\rho_f\,\rho_i}\,T_{fi},\quad\rho=\frac{k\,E\,\omega}{z}
\label{taut}
\ee
where $k\,(E,\omega)$ are the on-shell three momentum (baryon, meson energies) of the initial or final meson-baryon system. In order to extract the pole residue, we expand the amplitude $T^{(2)}$ on the second sheet in a Laurent series around the pole position, 
\be
T^{(2)\,i\to j}&=&\frac{a_{-1}^{i\to j}}{z-z_0}+a_0^{i\to j}+{\cal O}(z-z_0).
\label{pa}
\ee
The residue $a_{-1}$ and constant term $a_0$ can be obtained by a closed contour integration along a path $\Gamma (z)$ around the pole position $z_0$,
\be
a_n&=&\frac{1}{2\pi i}\oint_{\Gamma (z)} \frac{T^{(2)}(z)\,dz}{(z-z_0)^{n+1}}.
\label{contourint}
\ee
Alternatively, $a_{-1}$ can be expressed in terms of dressed quantities~\cite{Doring:2009bi},
\be
a_{-1}&=&\frac{\Gamma_D\,\Gamma_D^{(\dagger)}}{1-\frac{\partial}{\partial
    z}\Sigma}
\label{resasga}
\ee
where $\Gamma_D$ ($\Gamma_D^{(\dagger)}$, $\Sigma$) is the dressed
annihilation vertex (creation vertex, self-energy) as defined in Ref. \cite{Doring:2009bi}, evaluated on the second sheet at $z_0$. We have explicitly checked for all resonances, using the explicit values for $\Gamma_D, \,\Sigma$, that the results of Eqs. (\ref{contourint}) and (\ref{resasga}) agree.

The dressed quantities in Eq. (\ref{resasga}) are given in terms of the non-pole part $T^\npo$, which dresses the bare creation and annihilation vertices $\gamma^{(\dagger)}_B$, $\gamma_B$. For details see Ref. \cite{Doring:2009bi}. Note that for a simple energy and momentum independent $s$ wave interaction, $\gamma^{(\dagger)}_B=\gamma_B$ while for higher spins and partial waves the connection between bare annihilation and creation vertices can be more complicated.

Using Eqs. (\ref{taut}) and (\ref{pa}), the pole residues $R=|R|e^{i\theta}$ as quoted by the PDG~\cite{Amsler:2008zz} can be calculated. 
For the residue phase $\theta$~\cite{Amsler:2008zz} we consider the usual \cite{Hohler93} definition given by
\be
\tau=\tau_B+\frac{|R|\,e^{i\theta}}{M-z-i\Gamma/2}
\label{usualphi}
\ee
for a resonance with width $\Gamma$ on top of a background $\tau_B$. Comparing Eq. (\ref{usualphi}) with Eq. (\ref{pa}) and using Eq. (\ref{taut}), the pole residue $R$ and its phase are given by
\be
|R|&=&|a_{-1}\,\rho_{\pi N}|\non
\theta &=&-\pi+\arctan\left[\frac{{\rm Im}\,(a_{-1}\,\rho_{\pi N})}{{\rm Re}\,(a_{-1}\,\rho_{\pi N})}\right]
\label{rerere}
\ee
where $\rho_{\pi N}$ is the phase space factor $\rho$ from Eq. (\ref{taut}) for the $\pi N\to\pi N$ transition, evaluated at the complex pole position. 

Poles of the amplitude are searched for on the second sheet, as defined and described in Sec. \ref{sec:anal}. The results for pole positions and residues are summarized in Table \ref{tab:reso}. The extracted resonance parameters are compared with other studies \cite{Cutkosky:1979fy,Hohler93,Arndt:2006bf}, all of them accepted by the PDG \cite{Amsler:2008zz}. 

\begin{table}
\caption{Resonance parameters in the present study. The $z_0$ are the pole positions. The moduli $|R|$ and phases $\theta$ of the residues correspond 
to the $\pi N$ decay channel. For every resonance, the first line quotes the results of the present study, 
followed by the values from Refs. \cite{Arndt:2006bf,Hohler93,Cutkosky:1979fy} as quoted in the PDB \cite{Amsler:2008zz}. Resonances are included in the J\"ulich model via explicit $s$ channel exchanges except for the Roper resonance which appears dynamically generated.}
\begin{center}
\begin{tabular}{lllll}
 \hline\hline
\hspace*{0.cm}			&Re $z_0$\hspace*{0.cm}		&-2\,Im $z_0$\hspace*{0.cm}		&$|R|$\hspace*{0.cm}			&$\theta$ [deg]			\\
				&[MeV]				&[MeV]					&[MeV]					&[$^0$]				\\
$N^*(1440)\,P_{11}$ 		&1387	  			&147  	    				&48 	   				&-64 				\\
\cite{Arndt:2006bf}		&1359	  			&162		 			&38		 			&-98  				\\
\cite{Hohler93}			&1385	  			&164		 			&40		 			&				\\
\cite{Cutkosky:1979fy}		&1375$\pm 30$			&180$\pm 40$				&52$\pm 5$				&-100$\pm$35  			\\
$N^*(1520)\,D_{13}$ 		&1505	  			&95		    			&32 	   				&-18   				\\
\cite{Arndt:2006bf}		&1515	  			&113		 			&38		 			&-5  				\\
\cite{Hohler93}			&1510	  			&120		 			&32		 			&-8				\\
\cite{Cutkosky:1979fy}		&1510$\pm 5$			&114$\pm 10$				&35$\pm 2$				&-12$\pm$5  			\\
$N^*(1535)\,S_{11}$		&1519				&129		       			&31	       				&-3  				\\
\cite{Arndt:2006bf}		&1502	  			&95		 			&16		 			&-16  				\\
\cite{Hohler93}			&1487	  			&		 			&		 			&  				\\
\cite{Cutkosky:1979fy}		&1510$\pm 50$			&260$\pm 80$				&120$\pm 40$				&+15$\pm$45  			\\
$N^*(1650)\,S_{11}$ 		&1669				&136					&54					&-44				\\
\cite{Arndt:2006bf}		&1648	  			&80		 			&14		 			&-69  				\\
\cite{Hohler93}			&1670	  			&163		 			&39		 			&-37				\\
\cite{Cutkosky:1979fy}		&1640$\pm 20$			&150$\pm 30$				&60$\pm 10$				&-75$\pm$25  			\\
$N^*(1720)\,P_{13}$ 		&1663	  			&212  	    				&14 	   				&-82	   			\\
\cite{Arndt:2006bf}		&1666	  			&355		 			&25		 			&-94  				\\
\cite{Hohler93}			&1686	  			&187		 			&15		 			&				\\
\cite{Cutkosky:1979fy}		&1680$\pm 30$			&120$\pm 40$				&8$\pm 12$				&-160$\pm$30  			\\
$\Delta(1232)\,P_{33}$		&1218	  			&90		    			&47 					&-37	     			\\
\cite{Arndt:2006bf}		&1211	  			&99		 			&52		 			&-47 				\\
\cite{Hohler93}			&1209	  			&100		 			&50		 			&-48				\\
\cite{Cutkosky:1979fy}		&1210$\pm 1$			&100$\pm 2$				&53$\pm 2$				&-47$\pm$1  			\\
$\Delta^*(1620)\,S_{31}$ 	&1593	  			&72		    			&12 	   				&-108				\\
\cite{Arndt:2006bf}		&1595	  			&135		 			&15		 			&-92  				\\
\cite{Hohler93}			&1608	  			&116		 			&19		 			&-95				\\
\cite{Cutkosky:1979fy}		&1600$\pm 15$			&120$\pm 20$				&15$\pm 2$				&-110$\pm$20  			\\
$\Delta^*(1700)\,D_{33}$	&1637	  			&236  	    				&16 					&-38	     			\\
\cite{Arndt:2006bf}		&1632	  			&253		 			&18		 			&-40 				\\
\cite{Hohler93}			&1651	  			&159		 			&10		 			&				\\
\cite{Cutkosky:1979fy}		&1675$\pm 25$			&220$\pm 40$				&13$\pm 3$				&-20$\pm$25  			\\
$\Delta^*(1910)\,P_{31}$ 	&1840	  			&221  	    				&12 	   				&-153	   			\\
\cite{Arndt:2006bf}		&1771	  			&479		 			&45		 			&+172  				\\
\cite{Hohler93}			&1874	  			&283		 			&38		 			&				\\
\cite{Cutkosky:1979fy}		&1880$\pm 30$			&200$\pm 40$				&20$\pm 4$				&-90$\pm$30  			\\
\hline\hline
\end{tabular}
\end{center}
\label{tab:reso}
\end{table}

For prominent resonances with large branching to $\pi N$, the different analyses are in reasonable agreement with the present results. For other resonances that are wide and/or couple only weakly to $\pi N$, the results are much more disperse and there are noticeable differences among the results of Refs. \cite{Arndt:2006bf,Hohler93,Cutkosky:1979fy} and also to the results of the present study. Note that for resonances such as $\Delta^*(1910)$ and $N^*(1720)$, little is known about residues and phases. Given, e.g., the span of $\theta=+172^0,\,-90^0$ from the PDG for the $\Delta^*(1910)$, it is no surprise that the value of the present study of $\theta=-153^0$ does not match any of the two results. 

The Roper $N^*(1440)P_{11}$ resonance does not require a genuine pole term in the J\"ulich model \cite{Krehl:1999km}. Instead, the resonance shape is dynamically generated from the coupled channel interaction together with the unitarization from Eq. (\ref{bse}). Here, we can confirm this result, because we have indeed found a pole for this resonance [cf. also Sec. \ref{sec:roper}]. 

A special situation is given for the $S_{11}$ partial wave in which two resonance interfere making the extraction of resonance parameters more difficult. Table \ref{tab:reso} shows that the values from the PDB~\cite{Amsler:2008zz} for both $|R|$ and $\theta$ are very different in the various studies for the $N^*(1535)$. This is due to the systematic uncertainties from the close-by $\eta N$ threshold plus the interference with the $N^*(1650)$. The values of the present study lie within these wide spans. This issue is further discussed in Sec. \ref{sec:interference}.

\begin{figure}
\includegraphics[width=0.42\textwidth]{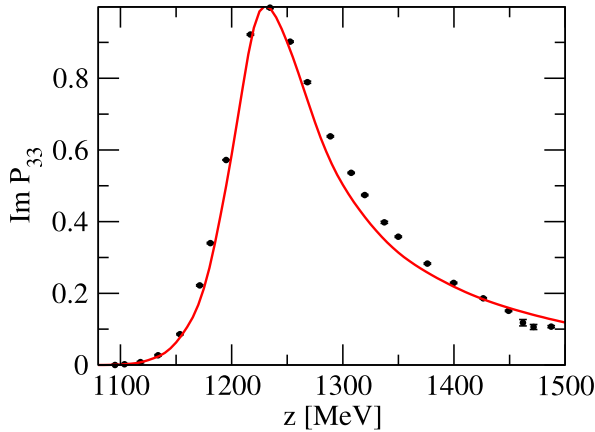}
\caption{Imaginary part of the $P_{33}$ amplitude with the $\Delta(1232)$. There are some residual deviations of the present fit (red solid line) and the SES [FA08] from Refs. \cite{Arndt:2008zz,Dugger:2009pn} (data points).}
\label{fig:p33im}
\end{figure}

For the other resonances, the results of the present study sometimes lie at the borders of the ranges quoted in the PDG. For the $\Delta(1232)$, the present values for pole position and residue lie even outside the range of the other studies. In Fig. \ref{fig:p33im}, the present solution is shown together with the SES solution of the partial wave analysis of Ref. \cite{Arndt:2008zz}. 
There are residual discrepancies for ${\rm Im}\,P_{33}$ which suggest that the J\"ulich model may need some fine-tuning to obtain an improved fit. It is a very small effect, but the PWA results are quite precise for the $\Delta(1232)$. Note, however, that even for near identical phase shifts $K$ matrix analyses and analyses using analytic $T$ matrices will give in general different pole positions and residues as a result of the different analytic properties.

Poles and zeros are important parameters and determine the global appearance of a partial wave amplitude. In the one-channel case, unitarity leads to zeros on the first sheet if there is a pole on the second sheet. For the coupled channel case discussed here, there is no such direct connection between poles and zeros.  Yet, zeros play an important role. For example, as pointed out in Ref. \cite{Gasparyan:2003fp}, the unitarity constraint in $\pi N$ to $\eta N$ leads to a zero above the $N^*(1650)$ in the absence of additional inelasticities; this results in an unobserved dip in the $\pi N\to\eta N$ cross section. The introduction of couplings of the $S_{11}$ resonances to $\pi\Delta$ in Ref. \cite{Gasparyan:2003fp} solved the problem, simultaneously reducing the maximum of the $\pi N\to\eta N$ cross section at the $N^*(1535)$ energies to the physical value.
	
\begin{table}
\caption{Position of zeros of the full amplitude $T$ in [MeV]. There is always another zero at the complex conjugate position according to Eq. (\ref{mirror}). For comparison, the zeros determined in Ref. \cite{Arndt:2003if} (FA02) are also quoted.}
\begin{center}
\begin{tabular}{lllll}
 \hline\hline
\multicolumn{2}{l}{first sheet}\hspace*{1cm}	& \multicolumn{2}{l}{second sheet}\hspace*{1cm}	& Ref. \cite{Arndt:2003if} \\
$P_{11}$	&$1235-0\,i$    &$S_{11}$	&$1587-45\,i$		&$1578-38\,i$		\\
$D_{33}$	&$1396-78\,i$   &$S_{31}$	&$1585-17\,i$		&$1580-36\,i$		\\
		&		&$P_{31}$	&$1848-83\,i$		&$1826-197\,i$		\\
		&		&$P_{13}$	&$1607-38\,i$		&$1585-51\,i$		\\
		&		&$P_{33}$	&$1702-64\,i$	 	&--			 \\
		&		&$D_{13}$	&$1702-64\,i$	 	&$1759-64\,i$  	 \\
\hline\hline
\end{tabular}
\end{center}
\label{tab:zeros}
\end{table}
The zeros of the J\"ulich model have been determined in the present study, with the results shown in Table \ref{tab:zeros}. 
There is a zero on the physical axis for the Roper channel. This reflects the peculiar phase shift of the $P_{11}$ partial wave as discussed in Ref. \cite{Gasparyan:2003fp}. For the $S_{31}$, $P_{31}$, and $P_{13}$ partial waves, the various zeros on the second sheet are in close vicinity to the respective resonance poles. The zero of $S_{11}$ lies in between the two $S_{11}$ resonances and will be further commented on in Sec. \ref{sec:interference}.

In Table \ref{tab:zeros}, also the zeros extracted from the FA02 solution of Ref. \cite{Arndt:2003if} are shown. In their sheet numbering, our sheet 2 is their sheet 1. The global pattern is similar; in $S_{11}$ the zero is in between the two resonances, in $S_{31}$, $P_{31}$, and $P_{13}$ the zeros are correlated with the respective resonance poles. Most interestingly, we find zeros in $P_{33}$ and $D_{13}$ at the  $\rho N$ branch point associated with the quasi-two-particle singularity at $z_{b_2}(\rho N)=1702-64\,i$ MeV.

\subsection{Pole structure of the Roper resonance}
\label{sec:roper}
The poles from Table \ref{tab:reso} all lie on the second sheet. We have also searched for poles on other sheets for some selected cases. E.g., in the partial wave analysis SP06 of Ref. \cite{Arndt:2006bf}, a pole of the Roper has been found at $z_0=1359-81\,i$ MeV on the second $\pi\Delta$ sheet (in their counting: first sheet), and another one at $z_0=1388-83\,i$ MeV on the third $\pi\Delta$ sheet (their counting: second). Also in the present study, we find a second pole of the Roper on the third $\pi\Delta$ sheet at $z_0=1387 - 71\,i$ MeV which is just a few MeV away from the pole on the second sheet quoted in Table \ref{tab:reso}. As discussed in Sec. \ref{sec:selectsheets} this is rather a replica of the pole on the second sheet, without physical implications, than a genuinely new structure; indeed, within the J\"ulich model the coupling strength of the Roper to the $\pi\Delta$ channel is moderate \cite{inprep}; a change of sheets does not change much the resonance properties in this special case. This is also reflected in the value of the residue of the additional Roper pole: $a_{-1}(2\text{nd}\,\pi\Delta)/a_{-1}(3\text{rd}\,\pi\Delta)=1.06-0.01\,i$ for the $\pi N$ residues of the poles on the second and third $\pi\Delta$ sheets.

The rather different pole positions and residues of the two Roper poles in Ref. \cite{Arndt:2006bf} indicate a larger coupling to the $\pi\Delta$ channel, and in this case the branch point $b_2$ plays an important role as stressed in Ref. \cite{Arndt:2006bf} which makes a simple Breit-Wigner parameterization of the Roper questionable~\cite{Arndt:2006bf}. 

In the J\"ulich model, the $\pi\pi N$ inelasticity of the Roper is rather given by the effective $\sigma N$ channel \cite{Schutz:1998jx,Krehl:1999km}. The early onset of inelasticity in the $P_{11}$ partial wave is naturally explained by the $s$ wave character of the $\sigma N$ coupling to the Roper~\cite{Schutz:1998jx,Krehl:1999km}. In contrast, $\pi\Delta$ couples in $p$ wave to the Roper, and the centrifugal barrier renders the contribution small at low energies. Thus, a large $\pi\Delta$ coupling to the Roper would be needed to provide the necessary inelasticity at low energies. Also, it has been shown in Ref. \cite{Krehl:1999km}, that the persistently large	 inelasticity at higher energies can be explained more naturally by a large $\sigma N$ coupling.

For completeness, let us mention a similar situation for the $\rho N$ sheet: on the third $\rho N$ sheet, there are poles in $P_{13}$ and $D_{33}$ situated a few MeV away from their counterparts on the second $\rho N$ sheet, associated with the $N^*(1720)$ and $\Delta^*(1700)$, respectively. Again, none of the two resonances couples strongly to $\rho N$ \cite{inprep}, and replicas of the poles on the second sheet appear on the third sheet. The $\rho N$ sheet in $D_{13}$, however, is special and will be discussed separately in Sec. \ref{sec:pol3rhon}.

%%%%%%%%%%%%%%%%%%%%%%%%%%%%%%%%%%%%%%%%%%%%%%%%%%%%%%%%%%%%%%%%%%%%%%%%%%%%%%%%%%

\subsection{Couplings}
\label{sec:gvsgamma}
It is convenient to express the residues $a_{-1}$ in terms of a few parameters $g$, given the $n^2$ different residues for the transitions within $n$ channels. It is possible to write, for the residues into the $\pi N$ and $\eta N$ channels,
\be
a_{-1}^{i\to j}=g_i\,g_j
\label{defg}
\ee
with a unique set of $g_i$ quoted below. The $g_i$ will be referred to as couplings in the following.
Note that in the determination of the $g_i$ there is an overall undetermined sign which we have fixed by choosing the real part of the coupling constant $g_{\pi N}$ positive. 

In Table \ref{tab:couchan} we list the coupling strength $g_i$ of the resonances to the $\pi N$ and $\eta N$ channels. The couplings to the effective $\pi\pi N$ channels will be published elsewhere~\cite{inprep}. They are important in the description of the $\pi N\to\pi\pi N$ reactions.

\begin{table}
\caption{Resonance couplings $g_i$ $[10^{-3}\,{\rm MeV}^{-1/2}]$ to the channels $\pi N$ and $\eta N$. }
\begin{center}
\begin{tabular}{lll}
 \hline\hline
\hspace*{2.4cm}			&$\pi N$	\hspace*{1.6cm}		&$\eta N$ \hspace*{2.2cm}\\
$N^*(1440)\,P_{11}$ 		& 		 $11.2-5.0i$   	&$-0.1+0.0i$\\
$N^*(1520)\,D_{13}$ 		& \hspace*{0.15cm}$8.4-0.8i$   	&\hspace*{0.2cm}$0.16-0.60i$\\
$N^*(1535)\,S_{11}$		& \hspace*{0.15cm}$8.1+0.5i$ 	&$11.9-2.3i$\\
$N^*(1650)\,S_{11}$ 		& \hspace*{0.15cm}$8.6-2.8i$  	&$-3.0+0.5i$\\
$N^*(1720)\,P_{13}$ 		& \hspace*{0.15cm}$3.7-2.6i$   	&$-7.7+5.5i$\\
$\Delta(1232)\,P_{33}$		&    		 $17.9-3.2i$	&$-     $\\
$\Delta^*(1620)\,S_{31}$ 	& \hspace*{0.15cm}$2.9-3.7i$  	&$-     $\\
$\Delta^*(1700)\,D_{33}$	& \hspace*{0.15cm}$4.9-1.0i$ 	&$-     $\\
$\Delta^*(1910)\,P_{31}$ 	& \hspace*{0.15cm}$1.2-3.5i$   	&$-     $\\
\hline\hline
\end{tabular}
\end{center}
\label{tab:couchan}
\end{table}
The allowed couplings $g_i$ in Table \ref{tab:couchan} are all non-zero, while only a few bare couplings are included \cite{Gasparyan:2003fp}. E.g., the bare $\eta N$ coupling of the $N^*(1650)$ is zero while the corresponding $g$ in Table \ref{tab:couchan} is finite. This is because the rescattering in the unitary coupled channel model renders the residue finite even when the bare couplings may be zero. 

In Ref. \cite{Doring:2009bi}, the bare and dressed vertices and the residue of the $\Delta(1232)$ have been evaluated. In Table \ref{tab:renovertex} we quote the corresponding results for other resonances. The vertices have been evaluated on the second sheet at the pole positions of the respective resonances.
\begin{table}
\caption{Bare and renormalized vertices $\gamma^C$ and $\Gamma^C$ in $[10^{-3}\,{\rm MeV}^{-1/2}]$ for some resonances in the $\pi N\to\pi N$ transition. The last two columns show the ratios defined in Eq. (\ref{ratios}).}
\begin{center}
\begin{tabular}{lllll}
 \hline\hline
\hspace*{1.8cm}			&$\gamma^C$ \hspace*{1.2cm} 	&$\Gamma^C$ \hspace*{1.2cm}	&r [\%]	\hspace*{0.0cm}	&r' [\%]\\
$N^*(1520)\,D_{13}$ 		&$6.4-0.6i$			&$13.2+1.2i$			&53			&61	\\
$N^*(1720)\,P_{13}$ 		&$-0.1+5.4i$			&$0.9+4.8i$			&24			&45	\\
$\Delta(1232)\,P_{33}$		&$1.3+13.0i$			&$-2.8+22.2i$			&45			&40	\\
$\Delta^*(1620)\,S_{31}$ 	&$0.1+14.3i$			&$5.0+5.7i$			&130			&66	\\
$\Delta^*(1700)\,D_{33}$	&$5.4-0.8i$			&$6.7+1.0i$			&33			&54	\\
$\Delta^*(1910)\,P_{31}$ 	&$9.4+0.3i$			&$1.9-3.2i$			&222			&22	\\
\hline\hline
\end{tabular}
\end{center}
\label{tab:renovertex}
\end{table}
The second and third columns show bare and dressed vertices, as defined in Ref. \cite{Doring:2009bi}. The other columns show the ratios
\be
r&=&|(\Gamma_D-\gamma_B)/\Gamma_D|,\non 
r'&=&|1-\sqrt{1-\Sigma'}|,
\label{ratios}
\ee
i.e. the relative differences of dressed vertex versus bare vertex, and coupling versus dressed vertex. The renormalization of the bare vertex (ratio $r$) is large in almost all cases. But also dressed vertices and couplings can be very different (ratio $r'$), due to the appearance of the term $1-\Sigma'$ in the denominator of the expression for the residue from Eq. (\ref{resasga}). The conclusions from this behavior are the same as in Ref. \cite{Doring:2009bi}: bare and dressed vertices are model dependent quantities; the non-pole $T$ matrix $T^\npo$ from the model dependent decomposition $T=T^\po+T^\npo$ enters in the calculation of $\Gamma_D$. In contrast, the couplings $g_i$ provide a meaningful expansion parameter of the amplitude around the pole of the resonance, independent of the amplitude decomposition into pole and non-pole parts.

%%%%%%%%%%%%%%%%%%%%%%%%%%%%%%%%%%%%%%%%%%%%%%%%%%%%%%%%%%%%%%%%%%%%%%%%%%%%%%%%%%

\subsection{Resonance interference in $S_{11}$}
\label{sec:interference}
As an example of the analytic structure we discuss the $S_{11}$ partial wave. This is of particular interest, because there are two resonances in this partial wave. For a systematic discussion of other partial waves and their analytic structure, in particular $P_{11}$ with the Roper resonance, see Ref. \cite{inprep}.

\begin{figure}
\includegraphics[width=0.48\textwidth]{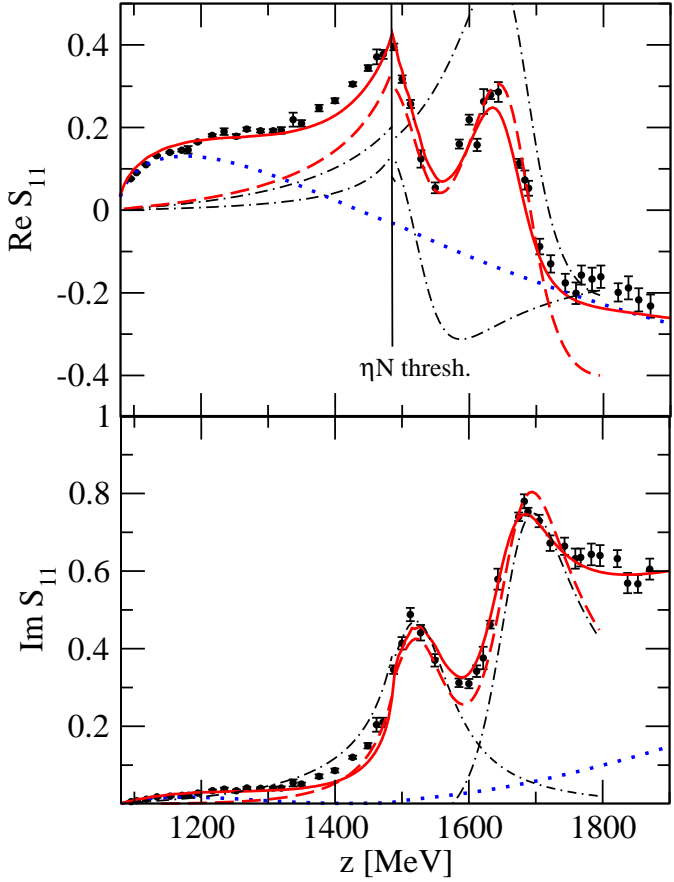}
\caption{Amplitude in the $S_{11}$ partial wave. The data points represent the single energy solution from the partial wave analysis [FA08] of Refs. \cite{Arndt:2008zz,Dugger:2009pn}. The solid red and dotted blue lines show the full amplitude and $T^\npo$, respectively. The dashed dotted lines show the pole approximations for the $N^*(1535)$ and the $N^*(1650)$ (from physical poles above the $\eta N$ threshold, from hidden poles below). The red dashed lines show their sum.}
\label{fig:s11}
\end{figure}
In Fig. \ref{fig:s11}, the amplitude $\tau$ for the $S_{11}$ partial wave is plotted, obtained from the amplitude $T$ via Eq. (\ref{taut}). The full solution of the J\"ulich model is indicated with the solid red lines. It describes well the SES solution of Ref. \cite{Arndt:2008zz} up to $z\sim 1.9$ GeV. The blue dotted lines indicate $T^\npo$. 
We can also plot the pole approximation from Eq. (\ref{pa}). For simplicity, we set $a_0=0$. On the physical axis, the pole approximations of the $N^*(1535)$ and $N^*(1650)$ appear as resonant like structures indicated with the black dashed-dotted lines. We first discuss the amplitude above the $\eta N$ threshold. For a discussion of the cusp and the amplitude below the $\eta N$ threshold, see below.

At first sight, the shapes of the two resonances in ${\rm Re}\,\tau$ are quite different: While the pole approximation of the $N^*(1535)$ shows a familiar shape with a maximum and a minimum in ${\rm Re}\,S_{11}$, the $N^*(1650)$ looks quite different. The reason is that $a_{-1}$ is a complex number that mixes real and imaginary parts of a classical Breit-Wigner shape. In other words, the phase of the resonances from Table \ref{tab:reso} is responsible for this twisting of resonance shapes and can have a very large effect.

The individual contributions from the two resonances (black dashed-dotted lines) are quite different from the full solution. However, the sum 
\be
T_a^{(2)}(z)=\frac{a_{-1}^{N^*(1535)}}{z-z_0^{N^*(1535)}}+\frac{a_{-1}^{N^*(1650)}}{z-z_0^{N^*(1650)}}, 
\label{tapps11}
\ee
indicated as the red dashed lines, fits the full solution quite well over the entire resonance region. Thus, the two resonances cannot be treated separately but must be treated together; the residue from the $N^*(1535)$ provides a strongly energy dependent background in the $N^*(1650)$ region and viceversa. Note, e.g. for ${\rm Re}\,\tau$ the strong energy dependence of the tail of the $N^*(1535)$ in the region of the $N^*(1650)$. The resonances interfere which each other. 

Thus, for a theoretical description one needs a unitary coupled channel model like the present one, which also allows for resonance interference. Otherwise, if one tries to extract resonance parameters individually for each resonance, one needs a substantial phenomenological background. Then, the parameters depend very strongly on that particular background and results are not reliable. 

At the level of pole positions and residues, resonance interference appears as a strong cancellation effect of pole approximations as discussed before; this should be clearly distinguished from what is conventionally meant with resonance interference: in a microscopic approach like the present one, the full partial wave amplitude with more than one explicit resonance can be decomposed according to
\be
T&=&T^\npo+T^\po\non
T^\po&=&\sum_{r,r'}\,(\Gamma_D)_r\,\frac{1}{(S_B^{-1})_r\,\delta_{rr'}-\Sigma_{rr'}}\,(\Gamma^{(\dagger)}_D)_{r'}
\label{resoint}
\ee
with resonance indices $r,r'$, dressed vertices $\Gamma_D$ and bare propagator $S_B$ as defined in Ref. \cite{Doring:2009bi} for the one-resonance case. Resonance interference is allowed by non-vanishing off-diagonal self energies $\Sigma_{12}$ and $\Sigma_{21}$. Second, even for $\Sigma_{12}=\Sigma_{21}=0$, the individual parts of the sum in Eq. (\ref{resoint}) can be large and make it difficult to phenomenologically disentangle the resonances, as in case of the $N^*(1535)$ and $N^*(1650)$ discussed here. 

Note that the full amplitude $T$ in Eq. (\ref{resoint}) is unitary which is automatically ensured by the complex phases of the dressed vertices $\Gamma$, that are linked to $T^\npo$ [cf. Ref. \cite{Doring:2009bi}]; the expression in Eq. (\ref{resoint}) for $T^\po$ should not be confused with a sum of two Breit-Wigner resonance amplitudes. While a single Breit-Wigner amplitude is unitary, a sum of two is not.

In contrast to the decomposition of Eq. (\ref{resoint}), the expansion of the amplitude in terms of poles and residues, as given in Eq. (\ref{tapps11}), does not rely on the model dependent separation into $T^\po$ and $T^\npo$~\cite{Doring:2009bi}; Eq. (\ref{tapps11}) allows to study the interference of resonances independent of the decomposition.

For the pole search it is important to find all poles, even if they couple only weakly to $\pi N$. For example, in Ref. \cite{Doring:2009bi} we have found poles in the so-called non-pole part of the amplitude $T^\npo$ (in $P_{33}$) which are dynamically generated and have no genuine pole term explicitly appearing in the amplitude. A useful way to identify poles and zeros is to consider the two curves defined by ${\rm Re}\,T^{(2)}(z)=0$ and ${\rm Im}\,T^{(2)}(z)=0$. In such a ``Gauss plot'', the curves intersect at poles and zeros. 

\begin{figure}
\includegraphics[width=0.4\textwidth]{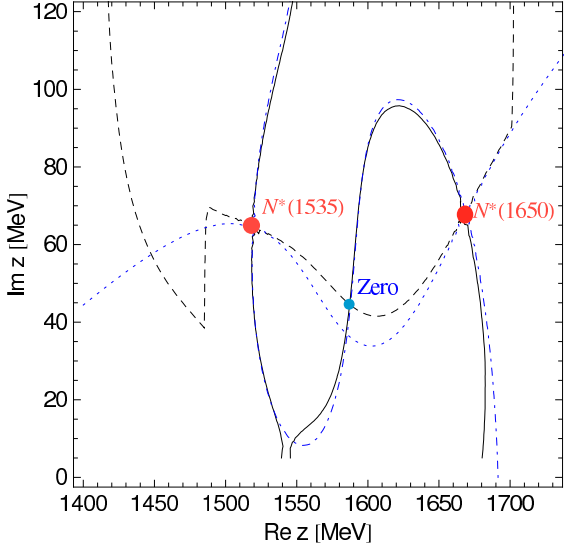}
\caption{The $S_{11}$ amplitude $T^{(2)}(z)$ on the second sheet. The lines are defined by ${\rm Re}\,T^{(2)}(z)=0$ and ${\rm Im}\,T^{(2)}(z)=0$. See text for further explanations.}
\label{fig:xray}
\end{figure}
In Fig. \ref{fig:xray} the analytic structure of the $S_{11}$ amplitude $T^{(2)}(z)$ on the second sheet is visualized in a Gauss plot. The solid line is defined by ${\rm Re}\,T^{(2)}(z)=0$, the dashed line by ${\rm Im}\,T^{(2)}(z)=0$. The two poles from Table \ref{tab:reso} and the zero from Table \ref{tab:zeros} are indicated in the figure and lie indeed at the intersections. There are no further intersections in the resonance regions.

The dashed dotted and dotted lines show the Gauss plot obtained from Eq. (\ref{tapps11}) instead of the full amplitude.
The sum from Eq. (\ref{tapps11}) reproduces remarkably well the amplitude even further away from the pole positions, including the position of the zero in between the two resonances. The kinks of the full solution (dashed line) at ${\rm Re}\,z=m_\eta+m_N$ and ${\rm Re}\,z={\rm Re}\,z_{b_2'}(\rho N)$ originate from the cuts appearing in the definition of the second sheet as discussed in Sec. \ref{sec:selectsheets}.

This comparison of full solution and the sum of Eq. (\ref{tapps11}), on the real axis and in the complex plane, reflects the accuracy of the extraction of pole positions and residues carried out in this study. There are, of course, still residual deviations between the full solution $T^{(2)}(z)$ and $T_a^{(2)}(z)$ from Eq. (\ref{tapps11}), e.g. for the position of the zero. They are due to higher order terms in the Laurent expansion which are, however, relatively small for the two $S_{11}$ resonances.

In the following, the amplitude below the $\eta N$ threshold is discussed. This part of the physical axis is directly connected to the part of the complex plane where the $\pi N$ channel is on the second sheet (2) but the $\eta N$ channel is on the first sheet (1). This sheet is called 21 in the following. The above discussed pole structure is not on sheet 21, but on sheet 22 (directly connected to the physical axis above the $\eta N$ threshold).

\begin{figure}
\includegraphics[width=0.48\textwidth]{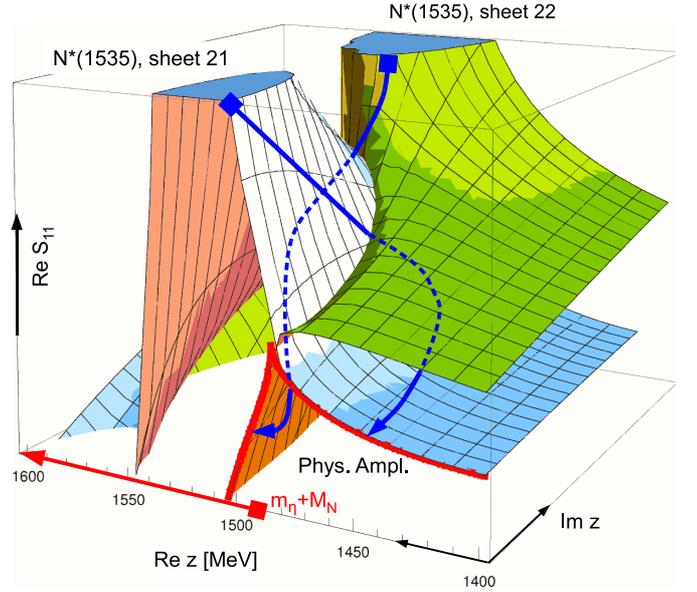}
\caption{Hidden poles in the $S_{11}$ amplitude. The $N^*(1535)$ and $N^*(1650)$ on sheet 22 ($\pi N,\,\eta N$ second sheet) are responsible for the resonant shapes above the $\eta N$ threshold, the hidden poles on sheet 21 ($\pi N$ second, $\eta N$ first sheet) are visible through their shoulders on the physical axis below the $\eta N$ threshold. The arrows indicate the impact of poles on different sheets to different pieces of the physical amplitude.}
\label{fig:hidden}
\end{figure}

Thus, the discussed pole structure on sheet 22 has no impact to the physical axis below the $\eta N$ threshold. However, on sheet 21 itself, there are also two poles. The real parts of their pole positions are greater than $m_\eta+m_N$. These ``hidden poles'' thus lie in the same $z$ region as the discussed $N^*(1535)$ and $N^*(1650)$, but on a different sheet. This situation is illustrated in Fig. \ref{fig:hidden}. The blue arrows indicate how the poles on different sheets are connected to the different pieces of the physical amplitude (red solid line). The figure shows schematically the different poles of the $N^*(1535)$; for the $N^*(1650)$, not shown in the figure, the situation is the same.

For the hidden poles on sheet 21, one can also draw the pole approximations and their sum from Eq. (\ref{tapps11}). This is indicated in Fig. \ref{fig:s11}, for $m_\pi+m_N<z<m_\eta+m_N$, with the black dashed dotted lines and the red dashed line.

As figure \ref{fig:s11} shows, the pole approximations from above the $\eta N$ threshold, in combination with the pole approximations from below the $\eta N$ threshold, provide a good reproduction of the pronounced $\eta N$ cusp (note $T^{NP}$ shows no cusp effect at all).

It is also worth mentioning that all resonance contributions become small as one approaches the $\pi N$ threshold (cf. dashed line in Fig. \ref{fig:s11}). This is a consequence of chiral symmetry which is included in the J\"ulich model through the modifications of Ref. \cite{Gasparyan:2003fp}. There, a derivative $\pi NN^*(1650)$ coupling provides the disappearance of resonance contributions close to threshold as required by chiral symmetry.

%%%%%%%%%%%%%%%%%%%%%%%%%%%%%%%%%%%%%%%%%%%%%%%%%%%%%%%%%%%%

\subsection{A pole on the third $\rho N$ sheet}
\label{sec:pol3rhon}
\begin{figure}
\includegraphics[width=0.4\textwidth]{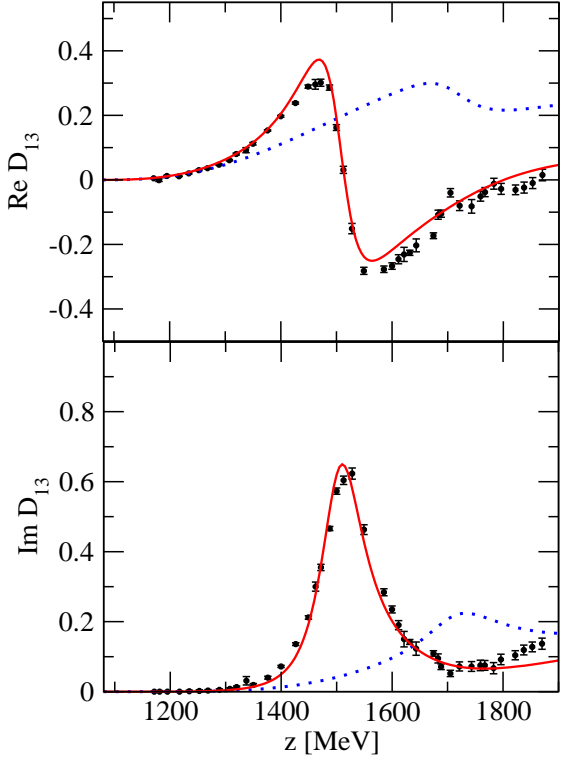}
\caption{Amplitude in the $D_{13}$ partial wave. Labeling of the plot as in Fig. \ref{fig:s11}.}
\label{fig:d13}
\end{figure}

The analytic properties of an amplitude with unstable particles can imply complex structures one of which is discussed in the following. 
In Fig. \ref{fig:d13}, the $D_{13}$ partial wave is shown. The full solution is indicated with the red solid lines and reproduces well the partial wave analysis from Ref. \cite{Arndt:2008zz}.

The blue dotted lines represent the non-pole part $T^\npo$ as defined in Ref. \cite{Doring:2009bi}, i.e. the amplitude without the $s$-channel resonance exchange diagrams. The pole term $T^\po$ contains the $s$ channel exchanges and the full amplitude is given by $T=T^\npo+T^\po$~\cite{Doring:2009bi}.  A resonant structure in $T^\npo$ is visible at around $z\sim 1.7$ GeV, which disappears in the full solution $T=T^\npo+T^\po$. The second Riemann sheet of $T^\npo$, however, is free of poles. Instead, the cut structure from the $\rho N$ branch point $b_2$ at $z_{b_2}=1702-64\,i$ MeV from Fig. \ref{fig:sn_structure} is enhanced. This is a sign that there is structure on the third $\rho N$ sheet [cf. Fig. \ref{fig:allfourslide}]. This is indeed the case. Using the prescriptions from Sec. \ref{sec:formalpath}, it is possible to analytically continue the amplitude of the J\"ulich model to the third $\rho N$ sheet. Indeed, there is a pole at $1613-83\,i$ MeV. It has a strong coupling to the $\rho N$ channel and a medium size coupling to $\pi N$ and is, thus, a state dynamically generated mainly from the attractive interaction in the $\rho N$ channel. 

Fig. \ref{fig:bgpoled13} illustrates the effects of a pole on the third sheet:
\begin{figure}
\includegraphics[width=0.451\textwidth]{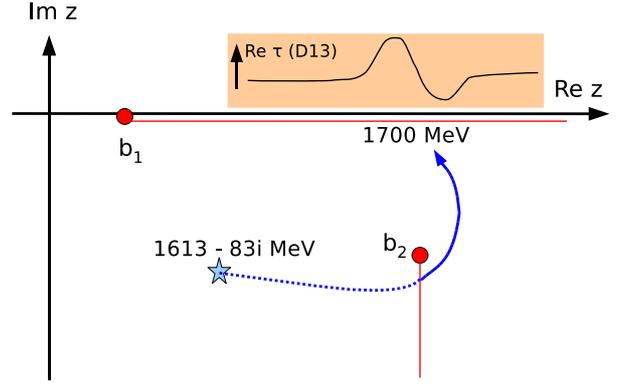}
\caption{The $D_{13}$ pole in $T^\npo$ on the third $\rho N$ sheet and its influence at the physical axis. The inset shows schematically the structure visible in $T^\npo$.}
\label{fig:bgpoled13}
\end{figure}
A path from the pole position to the physical axis must necessarily pass around the branch point $b_2$ to get from the third sheet to the second sheet that connects to the physical axis. The part of the path on the third sheet is indicated with the dotted lines in Fig. \ref{fig:bgpoled13}. The straight red line indicates our choice of the cut, where second and third sheet are connected [cf. Sec. \ref{sec:selectsheets}]. Once on the second sheet (solid blue line), it is possible to approach the physical axis on a straight path. 

In other words, the pole on the third sheet can affect the physical amplitude only via a detour around $b_2$. The structure in the physical amplitude is, thus, rather located at the position of $b_2$ than at the actual pole position. 

Summarizing, the branch points $b_2$ of effective $\pi\pi N$ channels lead to complex structures of the amplitude. The structure that appears as a resonance on the physical axis is in fact a threshold effect of the quasi-two-particle threshold $\rho N$ associated with $b_2$; it is induced by a pole at a different position, on a sheet that is not directly accessible from the physical axis. 

As a test, one can decrease the $\rho\pi\pi$ coupling of the $\rho$ self-energy. As a consequence, the $\rho$ becomes narrower and $b_2$ approaches the physical axis. Then, the structure that seemed to be a resonance in $T^\npo$ for the physical $\rho$, appears as a typical threshold cusp on the physical axis.

The mechanism discussed here may also apply to a very similar structure observed in Ref. \cite{Sarkar:2004sc}. There, $\Delta K$ scattering and its implication for the $\Theta^+$ pentaquark were discussed. For finite $\Delta$ width, a resonant structure appears close to the $\Delta K$ threshold which turns into a cusp structure once the $\Delta$ width is set to zero, in analogy to the present case; thus, what was noticed as a fading away of the pole \cite{Sarkar:2004sc} for some combinations of input parameters, may correspond to the pole moving far into the third sheet, in the present formulation.

Coming back to the discussion of $D_{13}$ in the J\"ulich model, in the full amplitude~\cite{Doring:2009bi} $T=T^P+T^{NP}$, the dynamically generated pole on the third sheet has disappeared due to a similar mechanism as discussed in Ref. \cite{Doring:2009bi} for the case of the $\Delta(1232)$; it is so far displaced from the physical axis that it has no visible effects any more.

The conclusions from Ref. \cite{Doring:2009bi} concerning poles in $T^\npo$ apply also for the $D_{13}$ partial wave: $T^\npo$ can be large and non-perturbative, associated with dynamically generated poles in the complex plane. However, poles in $T^\npo$ are systematically displaced far in the complex plane once $T^\po$ is added; only poles in the full $T$ matrix have physical significance. As a result, the $N^*(1520)$ appears as a clean resonance in $D_{13}$, with no additional structures in the amplitude.

%%%%%%%%%%%%%%%%%%%%%%%%%%%%%%%%%%%%%%%%%%%%%%%%%%%%%%%%%%%%
%%%%%%%%%%%%%%%%%%%%%%%%%%%%%%%%%%%%%%%%%%%%%%%%%%%%%%%%%%%%

\section{Summary and Conclusions}

The analytic properties of the $\pi N$ J\"ulich model have been determined. For this, a method for the analytic continuation for propagators with stable and unstable particles has been developed. Through the deformation of the integration contour it is possible to investigate all unphysical sheets. A channel with stable particles induces one branch point and two sheets for the amplitude. For effective $\pi\pi N$ channels, the pole of the unstable particle induces two additional branch points in the full propagator. In total, every effective $\pi\pi N$ channel induces three branch points and four sheets in the scattering amplitude. 

The pole positions and residues of the baryonic resonances up to a total spin of $J=3/2$ have been extracted. Residues, bare and dressed couplings have been compared, showing that only the residue provides a well-defined expansion parameter free of the model dependent decomposition into pole and non-pole part. For the $D_{13}$ partial wave, poles in the non-pole term $T^\npo$ are displaced far into the complex plane in the full amplitude $T=T^\po+T^\npo$; these findings are in line with the case of the $P_{33}$ partial wave discussed in a previous study. 

Resonance interference was shown to play a crucial role in $S_{11}$.
Taking this partial wave as an example, the quality of the extracted parameters has been shown; the residue terms alone provide already a good description of the full amplitude on the physical axis and also in the complex plane. 

The amplitudes of the J\"ulich model are derived within a field theoretical approach from Lagrangians obeying chiral constraints. Data are described to a high precision in the various partial waves. We claim that with these ingredients, in combination with a thorough treatment of the analytic properties, a reliable and precise extraction of pole positions and residues becomes possible. 

\vspace*{0.3cm}

\noindent {\bf Acknowledgements:} 
The work of M.D. is supported by DFG (Deutsche Forschungsgemeinschaft, Gz: DO 1302/1-1). This work is supported in part by the Helmholtz Association through funds provided to the virtual
institute ``Spin and Strong QCD'' (VH-VI-231), by the  EU-Research Infrastructure Integrating Activity
 ``Study of Strongly Interacting Matter" (HadronPhysics2, grant n. 227431)
under the Seventh Framework Program of EU and by the DFG (TR 16). F.H. is grateful to the support from the Alexander von Humboldt Foundation and the COSY FFE grant No. 41445282  (COSY-58).

%%%%%%%%%%%%%%%%%%%%%%%%%%%%%%%%%%%%%%%%%%%%%%%%%%%%%%%%%%%%
%%%%%%%%%%%%%%%%%%%%%%%%%%%%%%%%%%%%%%%%%%%%%%%%%%%%%%%%%%%%

\appendix
\section{The analytic structure of partial wave amplitudes}
\label{sec:app1}
It is well-known from dispersion theoretical considerations \cite{frauwa:1960,hoehlerpin} that below the $\pi N$ threshold there are various additional cuts and branch points. These cuts lead to analytic continuations of the scattering amplitude; however, due to their relatively large distance to the physical scattering region, poles on those sheets are rather of academic interest and have little influence on the partial waves. For completeness, in this Appendix the main features of the sub-threshold cuts are discussed. 

It is possible to identify the circular, short nucleon and other cuts with the partial wave projections of different diagrams of the present approach. First, there is the nucleon pole at $z=m_N$ in the $P_{11}$ partial wave, coming from nucleon $s$ channel exchange. In the current approach, the bare nucleon mass and bare $\pi NN$ coupling constant are renormalized by requiring the pole to be at $m_N$ and its residue to correspond to the physical $\pi NN$ coupling~\cite{Schutz:1998jx,Krehl:1999km}.

For $u$ and $t$ channel cuts, it is enough to consider the structure of $u$ and $t$ channel exchange processes contained in the present approach,  projected to partial waves via Legendre polynoms $P_\ell$. The two time orderings of the present TOPT formalism can be combined to structures appearing in the usual Feynman propagator according to 
\be
V_u&=&\int\limits_{-1}^1 dx\,\frac{P_\ell(x)}{u-m_N^2+i\epsilon}\non
V_t&=&\int\limits_{-1}^1 dx\,\frac{P_\ell}{t-m_t^2+i\epsilon}
\label{ut}
\ee
with $u$ channel nucleon exchange and $t$ channel exchange of particles with mass $m_t\geq 2m_\pi$ as given e.g. by the correlated $2\pi$ exchange in the $\sigma$ and $\rho$ channel, derived from $N\bar{N}\to\pi\pi$ pseudo-data via dispersion relations and using crossing symmetry~\cite{Schutz:1994ue,Schutz:1994wp}. Consider the on-shell to on-shell $\pi N\to\pi N$ kinematics as given by the transitions in Eq. (\ref{ut}).

\begin{figure}
\includegraphics[width=0.48\textwidth]{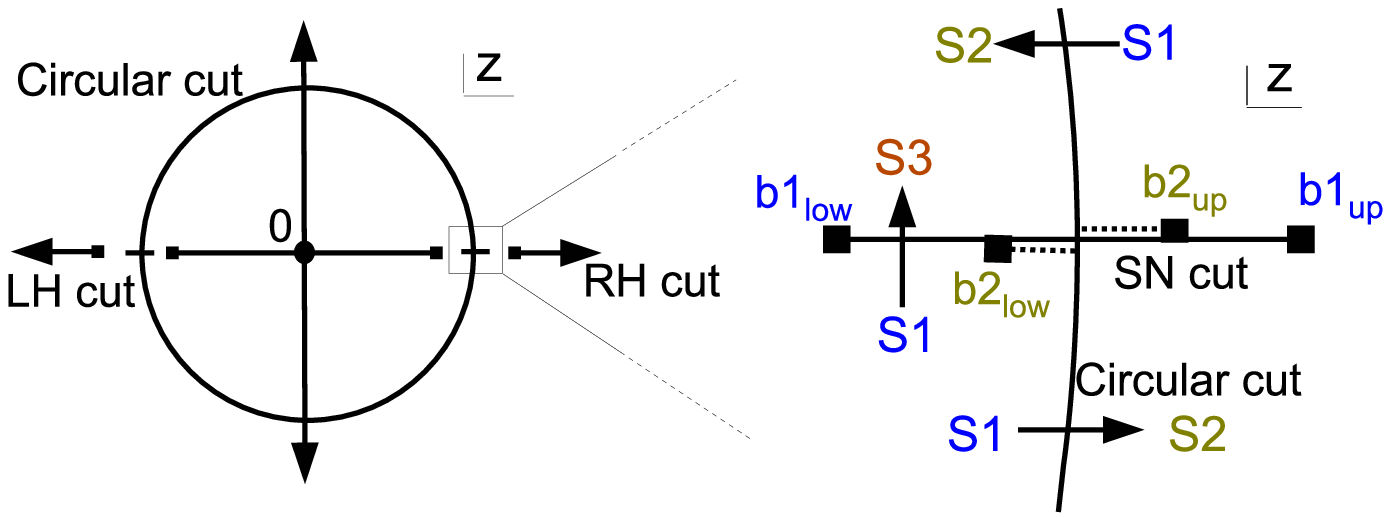}\\
\includegraphics[width=0.48\textwidth]{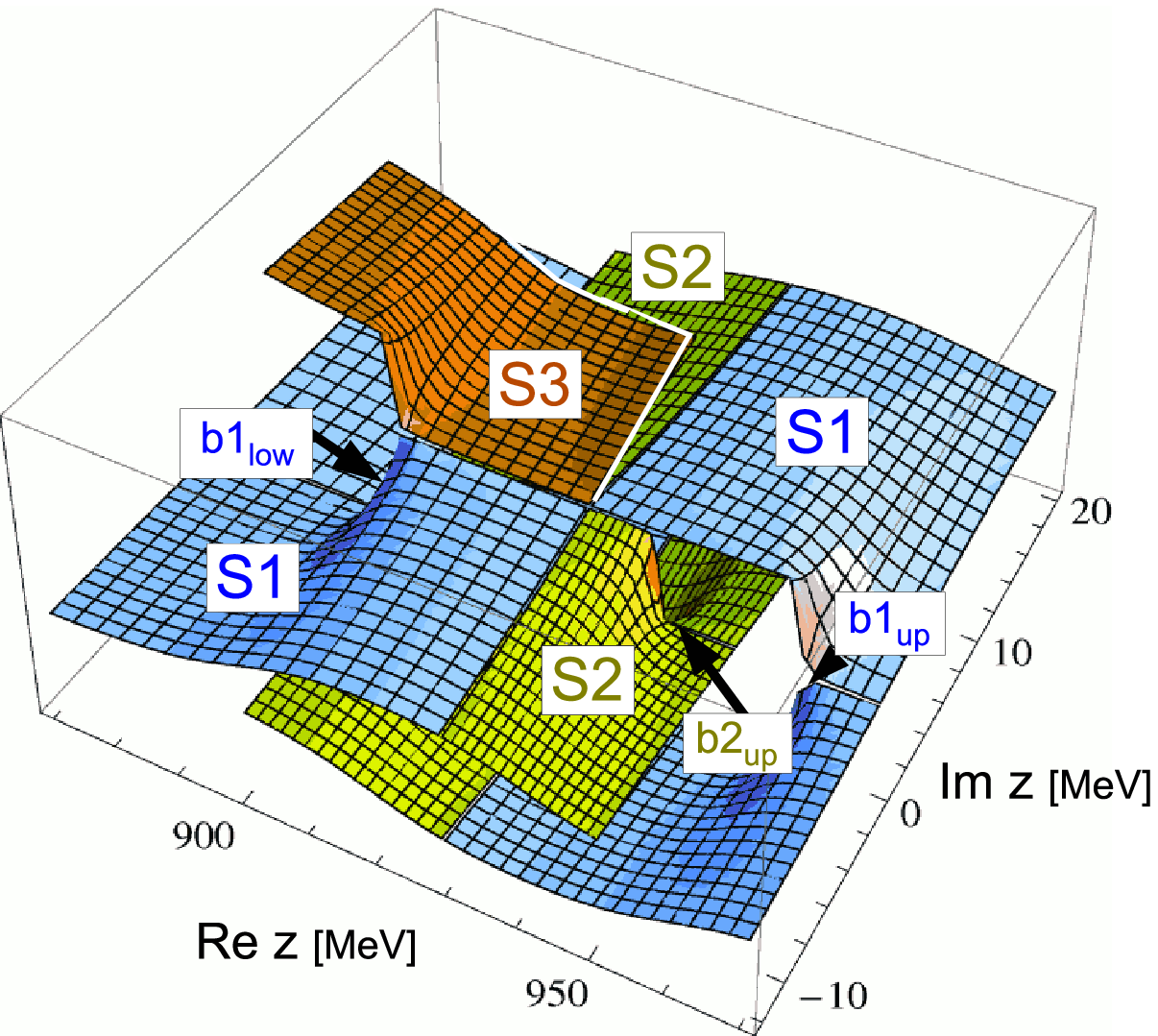}
\caption{The analytic structure of the $\pi N$ partial wave amplitudes below threshold. Upper left: right-hand (RH), left-hand (LH), circular and short nucleon cuts in the $z$-plane. Upper right: Analytic structure at the intersection of circular and short nucleon (SN) cut. The short nucleon cut (SN) starting at $b1_{{\rm low}}$ on sheet $S1$ ends at $b2_{{\rm up}}$ on sheet $S2$, whereas the short nucleon cut starting at $b2_{{\rm low}}$ on sheet S2 ends at $b1_{{\rm up}}$ on sheet $S1$.
Lower: The physical Riemann sheet is indicated with $S1$ (blue surface), the analytic continuations along the circular and short nucleon cuts are indicated with $S2$ (yellow surface) and $S3$ (orange surface), respectively. }
\label{fig:snc_cc}
\end{figure}
It is easy to see  that $V_u$ leads to the so-called short nucleon cut and both $V_u$ and $V_t$ contribute to the so-called circular cut. The branch points of the short nucleon cut can be obtained by solving for the zero of the denominator of $V_u$ at the borders of integration $x=\pm 1$,
\be
m_N^2-u|_{x=\pm 1}=0,\quad u=(p_\pi^f-p_N^i)^2
\ee
where $p_\pi^f$, $p_N^i$ are the final pion and initial nucleon momentum, respectively. The branch points of the short nucleon cut are situated above and below the $z=m_N$. There are two short nucleon cuts as indicated in the upper right of Fig. \ref{fig:snc_cc} due to the presence of the circular cut that crosses the short nucleon cut.

The projected potential $V_u$ can be analytically continued along the circular cut ($S1\leftrightarrow S2$), as indicated in the figure. The analytic continuation of the first Riemann sheet $S1$ along the circular cut (CC), situated at
\be
|z_{CC}|=\sqrt{m_N^2-m_\pi^2}\quad (929 \text{ MeV)}
\ee
is indicated as $S2$. The two sheets are given by
\be
V_u^{(S1)}&=&\int\limits_{-1}^1 dx\,\frac{P_\ell(x)}{(E_N-E_\pi)^2-2\,q_{{\rm on}}^2\,(1-x)-m_N^2}\non
V_u^{(S2)}&=&\int\limits_{-1}^1 dx\,\frac{P_\ell(x)}{(E_N+E_\pi)^2-2\,q_{{\rm on}}^2\,(1-x)-m_N^2}
\ee
with energies $E_i=\sqrt{q_{{\rm on}}^2+m_i^2}$ and the on-shell momentum $q_{{\rm on}}$ from Eq. (\ref{onstan}).
The short nucleon cut (SN) starting at $b1_{{\rm low}}$ on sheet $S1$ ends at $b2_{{\rm up}}$ on sheet $S2$, whereas the short nucleon cut starting at $b2_{{\rm low}}$ on sheet S2 ends at $b1_{{\rm up}}$ on sheet $S1$. The branch points are situated at 
\begin{align}
z(b1_{\rm low})	&=\frac{m_N^2-m_\pi^2}{\sqrt{m_N^2+2m_\pi^2}}&  \quad (899\text{ MeV, on S1)}\non
z(b2_{\rm up})	&=m_N					    &  	\quad (939 \text{ MeV, on S2)}\non
z(b2_{\rm low})	&=\frac{m_N^2-m_\pi^2}{m_N}		    &  	\quad(919 \text{ MeV, on S2)}\non
z(b1_{\rm up})	&=\sqrt{m_N^2+2m_\pi^2} 		    &  \quad(959 \text{ MeV, on S1)}.
\end{align}
Note the branch point at $z=m_N$ on the sheet $S2$ is not on the physical sheet $S1$, where the nucleon pole is situated~\footnote{The rather involved structure of the short nucleon cuts discussed here is not obvious in Ref. \cite{hoehlerpin}; in Ref. \cite{frauwa:1960} only one short nucleon cut is found.}.

The analytic continuation $S1\to S3$ and $S3\to S1$ along the short nucleon cut is given by (only $\ell=0$ is considered): 
\be
V_u^{(S3),\ell=0}=
\begin{cases}
V_u^{(S1),\ell=0}-\frac{\pi i}{q_{{\rm on}}^2}& \text{for Im }z>0\\
V_u^{(S1),\ell=0}+\frac{\pi i}{q_{{\rm on}}^2}& \text{for Im }z\le 0.
\end{cases}
\ee
for the part of the short nucleon cut with $z< |z_{CC}|$. This continuation is indicated with the orange surface in Fig. \ref{fig:snc_cc}. The branch points of the short nucleon cut are logarithmic, i.e. infinitely many sheets are connected at this branch point. See e.g. the surface $S3$ in Fig. \ref{fig:snc_cc} that connects to the next sheet (not drawn) below $z<b1_{\rm low}$. We do not quote the corresponding analytic continuations, nor the continuation along the short nucleon cut $S2\leftrightarrow S3$. 

As Fig. \ref{fig:snc_cc}, upper left, shows there is another short nucleon cut on the left-hand side obtained by $z\to-z$. There are additional cuts induced by $V_u$ from $0$ to $\pm i\,\infty$. Furthermore, there is an intermediate cut from $-m_N+m_\pi$ to $m_N-m_\pi$ coming from the kinematic factor $k$ appearing in Eq. (\ref{taut}), as well as a singularity at $z=0$.
Corresponding to the righthand, physical cut, there is the left-hand cut starting from $z=-m_N-m_\pi$ to $-\infty$. These cuts do not come from the partial wave projection but arise from the unitarity contained in Eq. (\ref{bse}) and symmetry $z\to -z$.
 
The $t$ channel exchange $V_t$ from Eq. (\ref{ut}) contributes to the circular cut and the cut from $0$ to $\pm i\,\infty$. While $u$ and $t$ channel exchange follow from crossing symmetry, there are many more interaction potentials contained in the present study. For example, the $\Delta(1232)$ $u$ channel exchange leads to a short $\Delta$ cut situated between $z=0$ and the short nucleon cut. The potentials contained in the present study~\cite{Krehl:1999km} lead to cut structures and branch points beyond the structures discussed in this Appendix.

Additionally, the exchange potentials appear not only in on-shell kinematics, as discussed here, but also in half-off-shell and off-shell kinematics. The three body cuts induced by such kinematics have to be carefully taken into account in the pole search far from the physical axis as discussed at the end of Sec. \ref{sec:implement}.

In this Appendix we have shown that the general analytic structure of the pion-nucleon scattering amplitude~\cite{frauwa:1960,hoehlerpin} can be identified with different ingredients of the J\"ulich model. In particular, all cuts and branch points demanded by crossing symmetry are present; circular and short nucleon cut, the intermediate cut from $-m_N+m_\pi$ to $m_N-m_\pi$, the cuts from $0$ to $\pm i\,\infty$, as well as the left-hand cut and a left-hand short nucleon cut obtained from $z\to-z$. 

These cuts and respective branch points can be indirectly visible, e.g. in a sharp rise of the Re $S_{11}$ amplitude close to the $\pi N$ threshold as shown in Fig. \ref{fig:s11}. Indeed, most of the amplitude close to threshold is given by nucleon exchange plus correlated two-pion exchange, as we have tested. In models without these potentials, the sharp rise at the $\pi N$ threshold can lead to the appearance of sub-threshold poles that mimic the missing cut structure~\cite{Arndtprivate}. Such a behavior has been recently found in Ref. \cite{Doring:2009uc}, where the meson baryon interaction is solely given by the Weinberg-Tomozawa contact term. Such sub-threshold poles should not be confused with genuine states.

%%%%%%%%%%%%%%%%%%%%%%%%%%%%%%%%%%%%%%%%%%%%%%%%%%%%%%%%%%%%%%%%%%%
%%%%%%%%%%%%%%%%%%%%%%%%%%%%%%%%%%%%%%%%%%%%%%%%%%%%%%%%%%%%%%%%%%%

\end{document}